\documentclass[10pt]{iopart}
\usepackage{iopams}
\usepackage{graphicx}

\begin{document}

\title[Fractional calculus within the optical model]{Fractional calculus within the optical model  used in nuclear and particle physics}

\author{Richard Herrmann}

\address{GigaHedron, Berliner Ring 80, D-63303 Dreieich, Germany}
\ead{herrmann@gigahedron.com}
\begin{abstract}
The optical model is a fundamental tool to describe scattering processes in nuclear physics.
The basic input is an optical model potential, which describes the refraction  and absorption processes 
more or less schematically.
Of special interest is the form of the absorption potential. With increasing energy of the incident projectile, a derivation of this potential must take into account the observed energy dependent transition from  surface to volume  type. The classic approach has weaknesses in this regard. 
We will discuss these deficiencies  and  
will propose an alternative method based on concepts developed within the framework of fractional calculus, which allows  to describe a smooth  transition from surface to volume absorption in an appropriate  way. 
\end{abstract}


%Uncomment for PACS numbers title message
%\pacs{00.00, 20.00, 42.10}
% Keywords required only for MST, PB, PMB, PM, JOA, JOB? 
%\vspace{2pc}
%\noindent{\it Keywords}: Article preparation, IOP journals
% Uncomment for Submitted to journal title message
%\submitto{\JPA}
% Comment out if separate title page not required
%\maketitle

\section{Introduction}
 In the development of physics the scattering experiments   by Geiger, Marsden and Rutherford \cite{gei09, rut11}  mark a twofold far-reaching highlight. 

A new view on the atom emerged, overcoming Thomson`s
plum pudding model \cite{tho04}  and splitting the atom into two distinct spatial areas, the outer atomic shell  and the inner nucleus.

This subsequently caused a specialization of research fields of the same kind, namely from a uniform view at the atom to two at first independent research areas of atomic shell physics, which in addition covers to a large extend the study of chemical reactions  and nuclear physics, which concentrates on the study of atomic nucleus itself. 

Consequently these experiments mark the beginning  of active research and systematic study  of the properties of nuclear matter.

Besides the passive observation of nuclear properties, which started with Becquerel`s \cite{bec96} accidental discovery  that uranium salts spontaneously emit a penetrating radiation, which he called  radioactivity and in the course of time unveiled spontaneous decay phenomena  (loosely sorted by emitted particle mass), like  $\gamma$,  $\beta$,  $\alpha$,
cluster decay \cite{ros84} even up to nuclear fission \cite{fle40} now the active scattering process allowed for a detailed study of the nucleus with different projectile-target combinations within a large range of incident energies.

Motivated by Joliot and Curie`s \cite{jol34}  experiments Fermi \cite{fer34} performed  a first systematic 
study  for different target nuclei  using slow neutrons \cite{fer54}  from a natural beryllium source  instead of 
$\alpha$-particles as projectiles bombarding  different chemical target elements.
Subsequent  experiments by Hahn and  Strassmann \cite{hah39} were correctly interpreted 
 as an induced fission of a nucleus by Meitner and Fritsch\cite{mei39}. 

The development of accelerators made it possible to realize scattering experiments with well defined projectile energies and different projectiles. While Hofstadter \cite{hof56} in Stanford used electrons to study the properties of nuclear matter, later light nuclei were used to produce trans-uranium and super-heavy elements \cite{oga04}. 

A milestone was the acceleration of uranium beyond the Coulomb-barrier and the studies on uranium-uranium scattering at  GSI in Darmstadt, which enabled  to generate  the strongest electromagnetic fields possible for a short time \cite{ale22}  and  to study the properties of large size nuclear molecules \cite{rei77, rei81}. 

Increasing the incident energy in heavy ion collisions allowed to investigate the properties of nuclear matter at extreme pressures and temperatures and thus to observe compression phenomena and possible phase transitions like deconfinement processes in quark-gluon-plasma \cite{raf82, raf86}. 

Last not least, the collision of two neutron stars \cite{opp39}  can be viewed as the ultimate nuclear scattering process  and may be a rare event; but it generates significant signatures though which may have  been  detected by the LIGO experiment \cite{abb17}.

In all these areas  a  vast amount of experimental scattering data for different projectile-target combinations over a large energy range has been accumulated in the last 120 years.  For an interpretation and categorization of these data appropriate theoretical models had to  be derived and applied. 

The non relativistic optical model was developed and  in the range from 1 - 200 MeV per nucleon  for the incident projectile it plays an outstanding role and has proven highly successful for analysing data on elastic and inelastic scattering of nucleons, deuterons and light elements \cite{kon14}.  

The basic idea behind the optical model is the 
interpretation of nuclear scattering using the  terminology of classical optics and interpreting nuclear matter as a nebular glass body.  For that purpose nuclear potentials are introduced such that the   elastic and inelastic contributions of a complex scattering process  are  described in terms of refraction and absorption processes. At first this is a purely phenomenological  ansatz which in the course of time  has been motivated by a derivation of the potentials used from reasonably chosen nucleus-nucleus interactions.

In this paper, we will concentrate on absorption processes and their energy dependence described with optical model potentials. With increasing energy of the incident projectile, a derivation of this potential must consider
 the observed smooth transition from surface to volume absorption. 
 
 The classic approach, which is widely accepted practice for more than 60 years now cannot take this into
 account appropriately. But what if there was a method that could do this?!

We will first discuss the classic approach and its deficiencies   and
we will then propose an alternative method, which 
allows to describe this transition adequately.

The new approach is based on concepts developed within the framework of fractional calculus and we will demonstrate its superiority. 

\section{The optical model - classical approach}
To begin with we will collect the necessary information on the classical derivation of optical model potentials as a basis for our criticism of the classical standard method.

The major idea behind the optical model is the representation of a nucleus by a mean field potential or optical potential 
$U(\vec{r}, E)$ being at least a
function of space coordinates and energy of the incident particle. 

The direct interaction of the incident particle  with a target nucleus is considered as an interaction 
with the optical potential only, leading to a quantum model and the corresponding Schr\"odinger equation
\begin{eqnarray}
\label{ctemp_se}
(- \frac{\hbar^2}{2 m} \Delta + U(\vec{r},E)) \Psi(\vec{r},E)  &=&  E \Psi(\vec{r},E)
\end{eqnarray} 
which is solved with appropriate boundary conditions leading to differential and total elastic  scattering angular distributions and the reaction cross section, which is equal to the sum of cross sections for all allowed inelastic processes. 

The optical potential is descibed by a complex quantity. Besides a real component $V(\vec{r},E)$ which accounts 
for elastic scattering only it also contains an imaginary component $W(\vec{r},E)$, which represents  all inelastic processes, happening  during the scattering process.
\begin{eqnarray}
\label{ctemp_potential}
U(\vec{r}, E) &=&  V(\vec{r},E) - i W(\vec{r},E)
\end{eqnarray} 

In view of the optical model,  these terms are interpreted as a description of refraction and absorption (note the minus sign)
processes during the scattering event respectively and were  first discussed as an appropriate approach for the nuclear scattering case by Ostrofsky and later Bethe \cite{ost36, bet40, hod67}.

Since it is found experimentally that scattered nucleons are polarised
even with an unpolarised incident beam, the optical
model potential is extended by a spin-orbit term 
\begin{eqnarray}
\label{ctemp_sopotential}
U_{so}(\vec{r}, E) &=&  ( V_{so}(\vec{r},E) - i W_{so}(\vec{r},E)) \vec{L} \cdot  \vec{s}
\end{eqnarray}
where $\vec{L}$ is the angular momentum and $\vec{s}$ are the Pauli spin operators. 
From experiment, there is no evidence for a significant imaginary spin-orbit contribution, so in general the $W_{so}$ term is ignored.

Finally, for protons as incident particles we have an 
additional Coulomb term $V_{c}(\vec{r})$.

The complete optical potential  therefore is given by:
\begin{eqnarray}
\label{ctemp_potentialtot}
U_{tot}(\vec{r}, E) &=&  V(\vec{r},E) - i W(\vec{r},E) + U_{so}(\vec{r}, E) + V_{c}(\vec{r})
\end{eqnarray}

There are two types of approaches, which historically lead step by step to a reasonable form 
of the optical potential:

First, a phenomenological approach, using analytical functions for well depths, e.g. Woods-Saxon potentials, where the parameters are adjusted using experimental data \cite{woo54, bec69}.

Second, we have microscopic optical potentials, which are based on an effective nucleon-nucleon interaction folded with reasonable nuclear density matter distributions, where in the idealized case there
is no parameter adjustment necessary\cite{sat79, var91, woo82, bau01a}.

A simplification, which is widely used in literature, is the restriction of the problem to spherical
symmetry. The Schr\"odinger equation ($\ref{ctemp_se}$) can be separated in spherical coordinates and we 
are left with the relevant radial part with central potentials $U_{tot}(\vec{r}, E)=U_{tot}(r, E)$,
with $r$ being the  distance from the scattering center and $E$ being the energy of the incident particle.  

For this case we may introduce the  form factors $f(r), g(r), h(r)$, which are  solely $r$-dependent: 
\begin{eqnarray}
\label{ctemp_potentialtot}
U_{tot}(r, E) &=&  V(E) f(r)  - i W(E) g(r)    + V_{so}( E) h(r)  + V_{c}(r)
\end{eqnarray} 
A classical choice for the central potential form factor $f(r)$ is given in close analogy to
nuclear density  or potential distributions used in shell model calculations or derived from 
mean field calculations in lowest order  is the Woods-Saxon potential \cite{woo54}:
\begin{equation}
\label{ctemp_f}
%f(r) =  \frac{1}{1 + e^{(r-R_0)/a_0}}
f(r) =  ( 1 + e^{(r-R_0)/a_0} )^{-1}
\end{equation} 
where $R_0$ is the average radius of the spherical target nucleus, which is given assuming incompressibility of nuclear matter as  $R_0 = r_0 A^{1/3}$ and $a_0$ is a measure of the skin size of the nucleus.

The absorption form factor $g(r)$ in (\ref{ctemp_potentialtot}) accounts for all inelastic processes. The standard narrative for the
energy dependence of this term reads as follows: 

For low energies, absorption
occurs mainly at the nuclear surface. Consequently for low energies the form of $g$ is chosen as the derivative of the 
Woods-Saxon potential, which defines a surface contribution $g( E \approx 0) = g_s$
\begin{equation}
\label{ctemp_gs}
g_s(r) =  -a_0 \frac{\partial}{\partial r} f(r)
\end{equation}
where $a_0$ with dimension length guarantees correct potential energy units. 
The spin-orbit form factor $h(r)$ in (\ref{ctemp_potentialtot}) is then given by the Thomas-form $g(r)/r$.
For higher energies absorption
 more and more happens throughout the nuclear volume. As a consequence being a function of energy a sliding transition from  surface- to volume-absorption is observed.

  Introducing an energy dependent  mixing coefficient $0 \leq \omega \leq 1$ and the two
  potential depths $W_s, W_v$ , with $W_s$  denominating the strength of the surface term and $W_v$ the strength of the volume term.  The imaginary part of the central potential is therefore given 
as:
\begin{eqnarray}
\label{linearAlpha}
W g(r) &=& \omega W_v  f(r) +  (1-\omega) W_s  g_s(r)   \\
  &=& \omega  W_v  f(r) - (1-\omega)   W_s  a_0 \frac{\partial}{\partial r} f(r)
\end{eqnarray}

 %--------------- figure frc
\begin{figure}[t]
\begin{center}
\includegraphics[width=\textwidth]{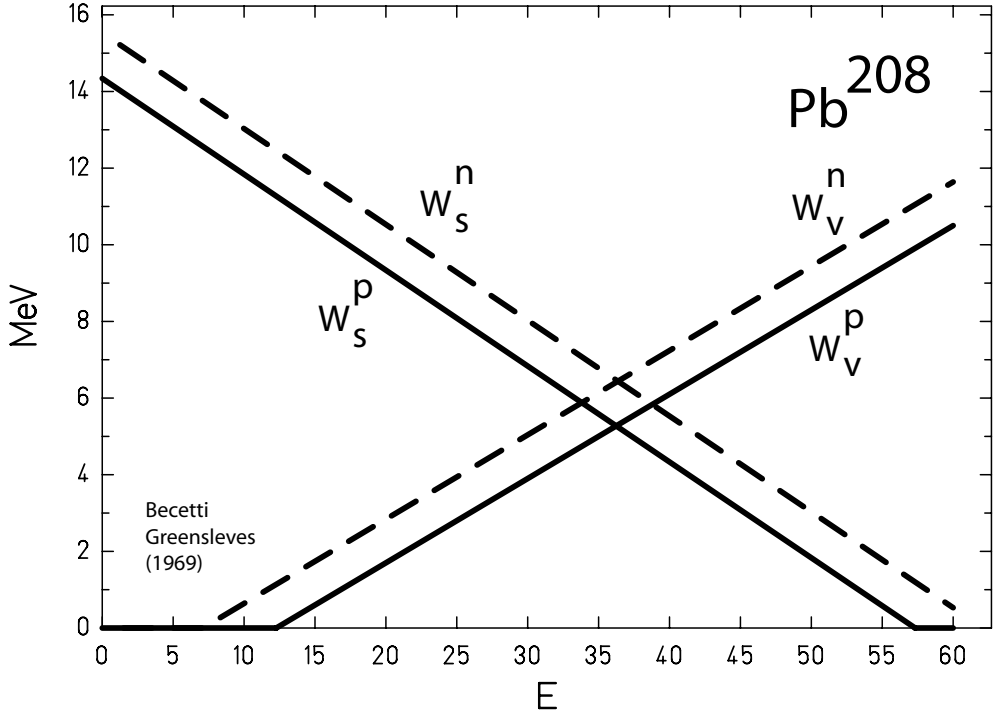}\\
\caption{
\label{wfig1}
{
Energy dependence for $W_v^I$ $W_s^I$ for $I \in \{ protons, neutrons\} $   for $^{208}\textrm{Pb}$ from 
\cite{bec69}. The physically valid area is given by $E< 50$ MeV.
} }
\end{center}
\end{figure}
%-----------------------------------------------------------------------

In other words: With increasing incident energy of the projectile the experimental data may be interpreted correctly assuming a smooth transition from surface to volume type absorption. 
Within the framework of optical potentials this behaviour is modelled by first introducing a surface potential defined as the rate of change of a given volume potential and subsequently  calculating  a weighted sum of these two
potentials. 

This procedure is common practice for more than 60 years now and has not been
seriously  questioned since its introduction. 
As an example, in figure \ref{wfig1} we have plotted the energy dependence of the factors 
$W_v$ and $W_s$ (the additional factor $4$ ensures comparable size range  of both factors)
\begin{eqnarray}
\label{parmsBec69}
 W g(r) &=& (W_v  - 4 a_0 W_s   \frac{\partial}{\partial r})  f(r) 
\end{eqnarray}
for increasing energy for $Pb^{208}$  using  parameters from Becetti and Greensleves \cite{bec69}, which are valid in the region 
$A > 40$ and $E < 50$ MeV and are given by:
\begin{eqnarray}
 W_v(E) &=&\max \left(0,  0.22 E -  \left\{
  \begin{array}{ll}
    2.70 & \textrm{protons}\\
    1.56 & \textrm{neutrons}\\
  \end{array}
\right. 
\right)\\
 W_s(E) &=& \max \left(0,  -0.25 E + 12.0 \frac{N-Z}{A} 
+  \left\{
  \begin{array}{ll}
    11.8 & \textrm{protons}\\
    13.0 & \textrm{neutrons}\\
  \end{array}
\right. 
\right) 
 \end{eqnarray}
 
We conclude, that a plausible physical concept has been realized at best only pragmatically, seemingly  good enough to 
serve as a tool in order to  classify the accumulated experimental data.  

A gradual  transition from surface to volume potential  is an essential requirement for a correct treatment of the energy dependence of the absorption term of  optical model potentials.  
This physical fact should be correctly  treated within the optical model, which has not been done so far.
Neither the physical justification for this requirement nor the mathematical treatment   has changed essentially through the last 60 years \cite{hod67, woo82, kon03}.
 
Already Roger Bacon in his \emph{opus maius} pointed out, that  one cause of error is the force of habit  \cite{bac67}. 

So it is time for a change and a chance for a more accurate description of scattering processes within 
the framework of the optical model.

In the following we will propose an  adequate treatment of the energy dependent absorption potential by using appropriate mathematical tools. 

Our approach follows a new  path  to generate  the progression between surface and volume absorption extending  the concept of a surface definition given in terms of standard differential vector calculus.

In the following we will  describe  a smooth transition between first derivative (surface) potential  and zeroth derivative, which means no  derivative (volume) potential  by applying  a fractional derivative of order $\alpha \in \mathbb{R}$ to the potential such that the absorption potential  becomes
\begin{eqnarray}
\label{ctemp_dfff}
W^\alpha  &=& W g(r) =  W_\alpha a_o^\alpha    \frac{\partial^\alpha}{\partial r^\alpha} f(r) \qquad  \quad 0 \leq \alpha \leq 1, \alpha \in \textrm{R}
\end{eqnarray}
For the cases $\alpha=0$ and $\alpha=1$ this reduces to the already known volume  and  surface type potential respectively, while
for all intermediate cases we obtain a new fractional  potential $W^{\alpha}$, which  
  is  much closer to the physical interpretation given above than the standard approach.
 
We will first present  the  surface definition based on classical vector calculus and then we will extend this framework   to the fractional case introducing an appropriate  fractional derivative and will apply new fractional vector calculus methods.

\section{The surface term in the optical model potentials}

Nucleon density $\rho(r)$ and collective volume potential $W_\textrm{vol}$ are related via a folding procedure of type 
\begin{equation}
\label{denVol}
W_\textrm{vol}(r) = W_v  \int_{G} d^3 r' w(r, r') \rho(r') = W_v (w \ast \rho)(r) =W_v  f(r)
\end{equation} 
with a weight $w(r,r')$, which models the effective short range nucleon-nucleon interaction in the collective model  
and the Coulomb interaction for protons respectively.
It should be emphasized, that a given potential may be the result of different weight/density combinations, e.g.  the weights  $w_{C}$  modelling  a Coulomb type interaction, which may be attractive, repulsive or zero 
for the interacting objects  with  $w_{\delta}$  being Dirac's $\delta$-function modelling an attractive strong contact interaction   and $w_{Y}$ being  a Yukawa type function respectively  modelling an
effective soft core interaction. 
\begin{eqnarray}
\label{wVol}
w_{C} &=&  \frac{e}{|r-r'|}  \qquad \qquad\qquad\qquad\qquad e \in \{ +1, -1, 0   \} \\
w_{\delta} &=& -\delta(|r-r'|) \\
w_{Y} &=& -e^{-\beta |r-r'|}/ |r-r'| \qquad\qquad\qquad \beta >0
\end{eqnarray} 
and densities
\begin{eqnarray}
\label{rhoVol}
%\rho_{ws} &=& \frac{1}{1 + e^{(r-R_0)/a_0}}\\
\rho_{ws} &=& (1 + e^{(r-R_0)/a_0})^{-1}\\
\rho_{H} &=&  H(|r-r'|)
\end{eqnarray} 
with $H$ being the Heavyside step function.

At first glance, it seems trivial, that a corresponding nuclear surface  potential $W_\textrm{surf}$
would be  directly related to the nuclear surface $S$
\begin{eqnarray}
\label{densur}
W_\textrm{surf}(r) &=& W_s  a_0 \int_{G} d^3 r' w(r, r') S(r') \nonumber \\
&=&  W_s  a_0 (w \ast S)(r) = W_s a_0 g_s(r) 
\end{eqnarray} 
with a scaling factor $a_0$ with dimension $[\textrm{fm}]$ in order to preserve potential energy units. 

But what is a nuclear surface?

An  elegant general definition of a surface, which actually is as appealing as Euclid's definition of a circle, defines  a surface via the spatial change of a density. In multi-dimensional space this change is calculated using the gradient operator $\nabla\cdot$ and this is how  differential 
calculus comes into play.  

Actually there are two candidates for a useful definition of a surface:

The first one generates an absolute value:

\begin{equation}
\label{s_inv}
S_{\textrm{magnitude}}(r) = |\nabla  \rho(r)| = \sqrt{\nabla  \rho(r)\cdot \nabla  \rho(r)}
\end{equation} 
This surface definition implies isotropy of the surface generation  since the gradient direction information is lost and only
the magnitude remains. It measures the maximum rate of density change at a given position $r$ and  is widely used within edge detection algorithms used in image processing \cite{gon18}.

The second reasonable definition is oriented:

\begin{equation}
\label{s_dir}
S_{\textrm{directional}}(r) =\vec{v} \cdot \nabla  \rho(r)  
\end{equation} 
We call this surface  directional since it is based on the definition of a directional derivative, which is given   by the projection of the density  gradient
on an arbitrary  vector $\vec{v}$ and gives the rate of density change in direction $\vec{v}$. Consequently 
in contrast to the $S_{\textrm{magnitude}}$ definition which always results in a positive sign for the surface, here  we obtain a positive sign for the surface for increasing density and a negative sign for decreasing density. 

In case of an optical model potential, the definition (\ref{s_dir}) seems appropriate, since the direction of 
the incident particle is important.

Therefore we obtain a possible definition for a surface potential  
\begin{eqnarray}
\label{denSur}
W_\textrm{surf}(r) &=& W_s a_0 \int_{G} d^3 r' w(r, r') \vec{v}' \cdot \nabla  \rho(r')
\end{eqnarray}

where the factor $a_0$ has dimension length to ensure correct energy units.
 
In order to make our argument as clear as possible, in the following we will discuss a simplified scenario;

In the following we restrict to spherically symmetric densities $\rho(r, \phi, \theta) = \rho(r)$,
 restrict to collective interactions $w(|r-r'|)$ where the spatial behaviour depends  on the distance  only,
 with the volume element $\sqrt g = r'^2  \sin(\theta')$
 and spherical surface shells setting
$\vec{v}(r,\phi,\theta) = \{-1,0,0\}$ (\ref{denSur}) simplifies to 
\begin{eqnarray}
\label{denVolR}
W_\textrm{Vol}(r) &=& W_v  \int_{G} \sqrt g  dr'  d\theta'  d\phi'   w(|r-r'|  \rho(r')\\
\label{denSurR}
W_\textrm{surf}(r) &=& -W_s a_0 \int_{G} \sqrt g  dr'  d\theta'  d\phi'   w(|r-r'| \frac{ \partial}{ \partial r'} \rho(r')
\end{eqnarray}

For the idealized case setting the collective nuclear interaction potential

 $w(|r-r')= -\frac{1}{4 \pi}\delta(|r-r'|)/\sqrt g  $we obtain the set of 
volume and surface potentials as    
\begin{eqnarray}
\label{denVolRR}
W_\textrm{vol}(r) &=& W_v  f(r)\\
\label{denSurRR}
W_\textrm{surf}(r) &=& -W_s a_0  \frac{ \partial}{ \partial r} f(r)
\end{eqnarray}
Thus the problem is reduced to the one dimensional case, which suffices to clarify our viewpoint. 

In order to model a gradual transition between these both limiting cases  (\ref{denVolRR}) and (\ref{denSurRR}) we will apply methods developed within the framework of fractional calculus 
\cite{old74, sam93, mil93, pod99,  hil00, mai10, ort11, her18}. 

The basic research area of  fractional calculus is to extend  the conceptual framework and the corresponding definitions of a derivative operator from integer order $n$ to arbitrary order $\alpha$, where $\alpha$ is a real or complex value or even more complicated a complex valued function  $\alpha=\alpha(r)$:
\begin{equation}
\label{c1first}
{d^n \over dr^n}   \rightarrow {d^\alpha \over dr^\alpha}     , 
\qquad\qquad\qquad\qquad\qquad n \in \mathbb{N}, \alpha \in \{\mathbb{R}, \mathbb{C}\}
\end{equation}
Several concepts coexist to realize this idea. In the following, we will first state the problem we want to solve and
will then present an appropriate solution, which will much better conform with the presented physical requirements.

\section{The optical model in view of fractional calculus}
Extending the concept of a derivative operator to  fractional order $\alpha$, where
$\alpha$ is a real number with the property $0\leq \alpha \leq 1$, such that we obtain a smooth transition between the cases $n=0$ and $n=1$ allows to extend the definition of an integer gradient to a fractional gradient operator too.  In cartesian coordinates we propose \cite{tar21}:
\begin{eqnarray}
\label{fracGradc}
\nabla^\alpha(x,y,z)  &=& (  \frac{ \partial^\alpha}{ \partial x^\alpha},
                                         \frac{ \partial^\alpha}{ \partial y^\alpha}, 
                                         \frac{ \partial^\alpha}{ \partial z^\alpha})  ,
 \quad \quad \quad \quad \quad    0\leq \alpha \leq 1, \alpha \in \mathbb{R}
\end{eqnarray}
or in spherical coordinates
\begin{eqnarray}
\label{fracGradsph}
\nabla^\alpha(r,\phi,\theta)  &=& (  \frac{ \partial^\alpha}{ \partial r^\alpha},
                                         \frac{1}{r} \frac{ \partial^\alpha}{ \partial \phi^\alpha}, 
                                         \frac{1}{r \sin(\theta)} \frac{ \partial^\alpha}{ \partial \theta^\alpha})  ,
 \quad 0\leq \alpha \leq 1, \alpha \in \mathbb{R}
\end{eqnarray}
and use this  fractional gradient  to define a unique fractional potential $W^{\alpha}$ in cartesian coordinates
\begin{eqnarray}
\label{denSuralpha}
W^{\alpha}(\vec{r}, \alpha) &=& W_\alpha a_0^\alpha \int_{G} d^3 r' w(r, r') \vec{v}' \cdot \nabla^\alpha  \rho(r')\\
&=&W_\alpha a_0^\alpha (w *  \nabla^\alpha \rho)(\vec{r}) \nonumber \\
&& \qquad\qquad 0\leq \alpha \leq 1, \alpha \in \mathbb{R}
\end{eqnarray} 
where the factor $a_0^\alpha$ ensures correct units and the potential depth $W_\alpha$ is now a function of $\alpha$. The limiting cases  corresponding to (\ref{denVolRR}) and (\ref{denSurRR}) are then 
\begin{eqnarray}
\label{denSuralpha0}
W^{\alpha}(r,\alpha = 0) &=& W_\textrm{vol}(r)\\
W^{\alpha}(r,\alpha = 1) &=& W_\textrm{surf}(r)
\end{eqnarray}
and consequently the limiting cases for the potential depths
\begin{eqnarray}
\label{denWralpha0}
W_\alpha(\alpha = 0) &=& W_v\\
W_\alpha(\alpha = 1) &=& W_s
\end{eqnarray}

For central potentials we may switch to spherical coordinates and with $\vec{v}(r,\phi,\theta) = (-1,0,0)$
and are lead to the fractional extension of (\ref{denSurRR}):
\begin{eqnarray}
\label{denSurRRx}
W^{\alpha}(r, \alpha) &=&  W_\alpha a_0^\alpha  \int_{G} \sqrt g  dr'  d\theta'  d\phi'   w(|r-r'| )
\frac{ \partial^\alpha}{ \partial r^{`\alpha}} \rho(r') \\
&=&W_\alpha a_0^\alpha (w *  \partial^\alpha \rho)(r) \qquad\qquad 0\leq \alpha \leq 1, \alpha \in \mathbb{R}
\end{eqnarray}
In the case of the attractive contact potential $w(|r-r'|)= -\frac{1}{4 \pi}\delta(|r-r'|)/\sqrt g  $ we explicitely obtain
for the spherically symmetric  fractional  absorption potential $W^{\alpha}$:
\begin{eqnarray}
\label{denradalpha}
W^{\alpha}(r,\alpha) &=& W_\alpha a_0^\alpha \frac{ \partial^\alpha}{ \partial r^\alpha} \rho(r), \qquad\qquad 0\leq \alpha \leq 1, \alpha \in \mathbb{R}
\end{eqnarray}
With (\ref{denSuralpha}) for the general cartesian and (\ref{denSurRRx}) for the spherically symmetric case respectively we propose fractional optical model potentials  for an adequate description of the energy dependence of nuclear absorption processes. In the following we will give closed form solutions for the 
important case of spherical Woods-Saxon type densities.

\section{Derivation and applications of the spherical  fractional model potential}
We interpret the scattering process as a mapping from an initial to final quantum states covering the full region $\{-\infty, +\infty\}$, which for the spherically symmetric case determines the bounds for the
radial component $0 \leq r < \infty$ and consequently determines the bounds for the fractional derivative 
definition.  

For an arbitrary density $\rho(r)$ we choose a  Liouville type fractional derivative \cite{lio32}  which is defined as a sequential operation: A fractional integral with bounds  $\{\infty, 0\}$  is followed by an ordinary derivative. 
\begin{eqnarray}
\label{dderivfrac}
\frac{ \partial^\alpha}{ \partial r^\alpha}\rho(r)   &=& \rho^{(\alpha)}(r) = \frac{ \partial}{ \partial r} I^{1-\alpha}(r) \rho(r)
\end{eqnarray}
Consequently, if we know the fractional integral $I^\alpha$, we also know the fractional derivative of $\rho(r)$.

The fractional integral $I^{1-\alpha} \rho(r)$ is given by a  convolution with a weakly singular kernel $w_L(h) = h^{-\alpha} $ and thus we will apply the following definition:
\begin{eqnarray}
\rho^{(\alpha)}(r)   &=&
\frac{ \partial}{ \partial r}
 (w_L \ast \rho)(r) \\
\label{dintfrac}
 &=&\frac{ 1}{ \Gamma(1-\alpha)} 
\frac{ \partial}{ \partial r}
\int_\infty^0   dh  \,  h^{-\alpha} \rho(r+h)   
\end{eqnarray}
This at first abstract fractional derivative definition (\ref{dintfrac})  allows a provisional  physical interpretation:

We rewrite  (\ref{dintfrac}) as a sum 
\begin{eqnarray}
\label{dderivfractraj}
\rho^{(\alpha)} &=&\frac{1}{2} \frac{ 1}{ \Gamma(1-\alpha)}  \frac{ \partial}{ \partial r}  
 \bigl( 
	 \int_{\infty}^0  \!\! dh  \,  h^{-\alpha} \rho(r-h)  -
	 \int_0^\infty      \!\! dh  \,   h^{-\alpha}  \rho(r+h) 
\bigr)  
\end{eqnarray}
Within a classical picture \cite {ehr27}, the fractional derivative is a weighted energy dependent sum of the projection of the density change  along the classical trajectory expectation value of an incoming projectile, which runs from $\infty \leq x \leq +\infty $  onto the radial vector. The weight function may be considered  as the idealized hadronic analogue to a Bragg energy deposition curve, with a singularity at the position of closest approach to the origin \cite{bra04, bra05, wil46}. 

Note that due to  our definition of a fractional derivative it follows a weak correspondence  for $\alpha$ being even and odd respectively, if analytically continued from $0 < \alpha < 1$ to $\alpha> 0$:
\begin{equation}
\label{dennorm4}
\rho^{(\alpha)}(r) = 
\left\{
\begin{array}{ll}
+\rho^{(n)}(r)                 &\quad \alpha  \rightarrow n, n \,\,\textrm{even} \\
-\rho^{(n)}(r)                &\quad \alpha  \rightarrow n, n  \,\,\textrm{odd} 
\end{array}
\right.
\end{equation}
which in a natural way  yields the required sign change in  (\ref{ctemp_gs}),  (\ref{denSurR})  and (\ref{denSurRR}) for volume and surface part respectively.

Let us now apply  (\ref{dintfrac}) to  the important case of a density of Woods-Saxon type
\begin{equation}
\label{ctemp_fn}
%\rho_{WS}(r) = \rho_0 \frac{1}{1 + e^{\frac{r-R_0}{a_0}}}
\rho_{WS}(r) = \rho_0 (1 + e^{\frac{r-R_0}{a_0}})^{-1}
\end{equation} 
we obtain an analytic solution for the fractional derivative of $\rho_{WS}$  according to (\ref{dderivfrac}):
\begin{eqnarray}
\label{dderivfrac2}
\rho^{(\alpha)}_{WS}(r) 
&=& \frac{ 1}{ \Gamma(1-\alpha)} \frac{ \partial}{ \partial r} \int_\infty^0  dh  \,  h^{-\alpha} \rho_{WS}(r+h), \qquad 0< \alpha <1 \\
&=&
-\rho_0 a_0^{1-\alpha}  \frac{ \partial}{ \partial r}   \textrm{Li}_{1-\alpha}(-e^{(R_0-r)/a_0})\\
&=&- \rho_0 a_0^{-\alpha}  \,  \textrm{Li}_{-\alpha}(-e^{(R_0-r)/a_0})\\
&=& \rho_0 a_0^{-\alpha}  \,  \textrm{F}_{-\alpha-1}((R_0-r)/a_0)
\end{eqnarray}
where $\textrm{Li}_\gamma(x)$ is the polylogarithm of fractional order $\gamma$, in our case  with $-1  < \gamma < 0$ or equivalently from a physicist's point of view $\textrm{F}_\beta(x)$ is the Fermi-Dirac integral of fractional order $\beta$, in our case with $-2 < \beta < -1$,   which occurred in physics at first for the special case $\beta = 1/2$ in the description of a degenerated electron gas in metals and later also within the framework of renormalization of quantum field theories \cite{pau27, som28, bla82, sac93}.

As a side note, since the Fermi-Dirac integrals are given as
\begin{eqnarray}
\label{fdintegrals}
F_\beta(r)&=& \frac{ 1}{ \Gamma(1-\beta)} \int_0^\infty  dh  \,  h^{\beta} 
(1 + e^{h-r})^{-1} , \qquad  \qquad  \beta >-1
\end{eqnarray}
and  the derivative follows from 
\begin{eqnarray}
\label{fdintegrals2}
F_{\beta(r)-1}(r) &=&   \frac{ \partial}{ \partial r}   F_{\beta}(r) 
\end{eqnarray}
we may explicitely note:
\begin{eqnarray}
\label{fdintegral32}
F_{-\alpha-1}(r) &=&    \frac{ 1}{ \Gamma(1-\alpha)}  \frac{ \partial}{ \partial r}  \int_0^\infty  dh  \,  h^{-\alpha} 
(1 + e^{h-r})^{-1} , \quad  0 \leq \alpha \leq 1
\end{eqnarray}
Comparison with (\ref{dderivfrac}), (\ref{dintfrac}), (\ref{dderivfrac2}) leads to the conclusion: Everybody who uses fractional  ($\alpha \notin Z$) Fermi-Dirac or Bose-Einstein integrals 
is actually doing fractional calculus \cite{nis12, mil96, din57, cha07}!

Let us summarize our results obtained so far:

We have derived a closed formula for the fractional derivative 
of a Woods-Saxon type density  in terms of fractional polylogarithms \cite{cos07, tao22}, which allows for a
smooth transition between the two extremal cases of volume and surface interpretation respectively. 

Normalization of the density $\rho^{(\alpha)}_{WS}(r) $ may be achieved by integrating the density to obtain the correct number  of protons/neutrons for a given nucleus. This integral may also be interpreted as a 
special case of a Mellin-transform \cite{mel97} of a Fermi-Dirac integral which has been  considered  by Dingle \cite{din57}. 
\begin{eqnarray}
\label{dennorm2}
\textrm{Vol}_{WS}(r,\alpha) &=& 4 \pi \int_o^\infty r^2 dr   \rho_{WS}^{(\alpha)}(r)\\
&=& 4 \pi \int_o^\infty r^2 dr   a_0^{-\alpha}\textrm{Li}_{-\alpha}(-e^{(R_0-r)/a_0}) \\
 &=&  8 \pi   a_0^{3-\alpha}\textrm{Li}_{3-\alpha}(-e^{(R_0-r)/a_0}) 
\end{eqnarray}

With the help of the asymptotic formula \cite{woo92}
\begin{eqnarray}
\lim_{w \rightarrow \infty}  \textrm{Li}_{p}(\pm e^w) &=&
-\frac{w^p}{\Gamma(1+p)}       \qquad\qquad    p \neq -1,-2,...,-n
\end{eqnarray}
in the limit ${a_0 \rightarrow \infty}$ it follows for a homogeneous sphere with radius $R_0$:
\begin{eqnarray}
\label{dennorm3}
\textrm{Vol}_{WS}(r,\alpha)  
 &=&  \lim_{a_0 \rightarrow \infty} 8 \pi   a_0^{3-\alpha}\textrm{Li}_{3-\alpha}(-e^{(R_0-r)/a_0}) \\
 &=&  \frac{8  \pi}{\Gamma(4-\alpha)}  R_0^{3-\alpha} 
\end{eqnarray}
which indeed yields the two limiting cases for volume and surface term normalization respectively:
\begin{equation}
\label{dennorm4}
\textrm{Vol}_{WS}(r,\alpha) =
\left\{
\begin{array}{ll}
\frac{4}{3} \pi R_0^3  &\quad \alpha = 0 \\
2 \pi R_0^2                &\quad \alpha = 1 
\end{array}
\right.
\end{equation}
Let us finally connect the fractional derivative density $\rho^{(\alpha)}_{WS}(r) $  with the fractional optical potential absorption term $V$ according to  (\ref{denSurRRx}):
\begin{eqnarray}
\label{Vderiva}
W^{\alpha}(r, \alpha) &=&  W_\alpha a_0^\alpha  \int_{G} \sqrt g  dr'  d\theta'  d\phi'   w(|r-r'| )
\frac{ \partial^\alpha}{ \partial r^{`\alpha}} \rho(r') \\
&=&W_\alpha a_0^\alpha (w \ast  \rho^{(\alpha)})(r) \\
&=&W_\alpha a_0^\alpha (w  \ast  w_L  \ast \rho)(r) 
\qquad\qquad 0\leq \alpha \leq 1, \alpha \in \mathbb{R}
\end{eqnarray}
with only one parameter potential depth $W_\alpha(\alpha)$.

  %--------------- figure frc
\begin{figure}[t]
\begin{center}
\includegraphics[width=\textwidth]{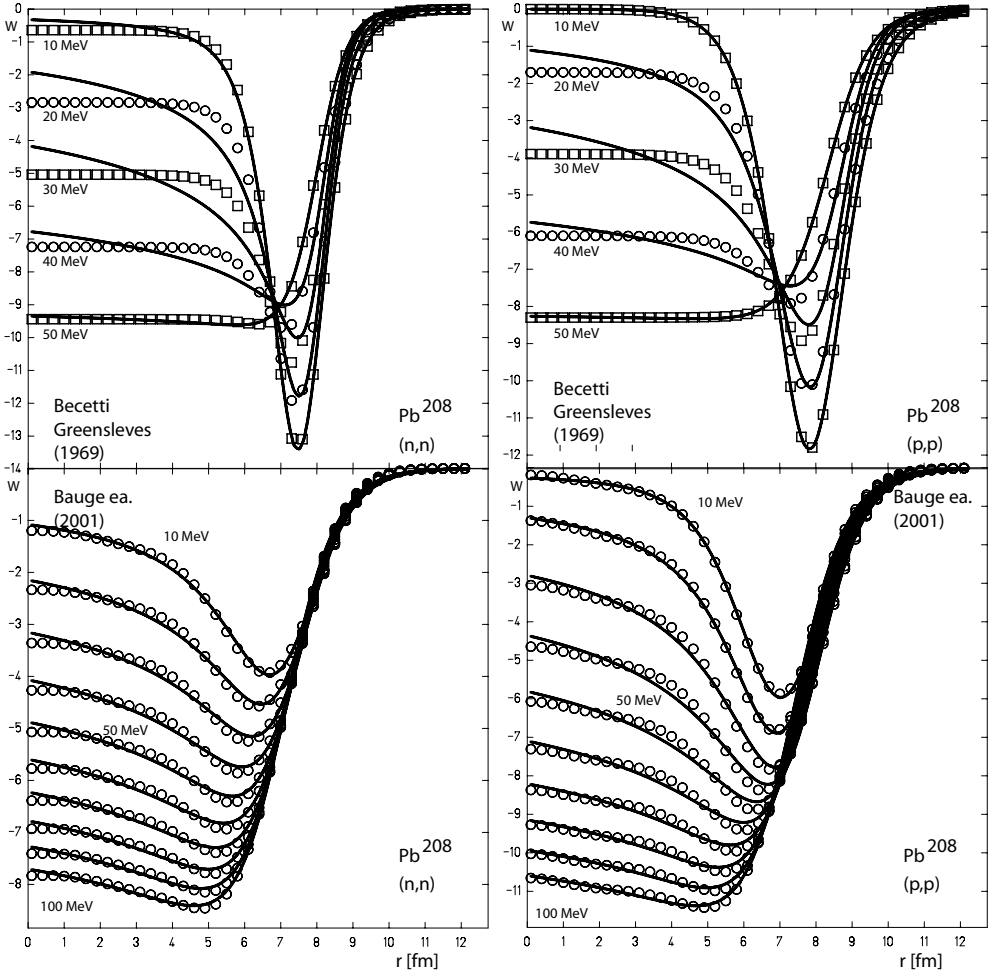}\\
\caption{
\label{wfig2}
{
Top: Using parameters (\ref{parmsBec69}) from Becetti and Greensleves the absorption potentials for incident neutrons (left side)  and protons (right side) respectively are plotted  in the energy range $10-50$ MeV for   \textrm{Pb}$^{208}$ (cirlces ans squares).  Solid lines show
the optimum fit in the range $0 < r < 2 R_0$ for the proposed fractional derivative based potential (\ref{fcWS}).   

Bottom:
Using parameters  derived from the microscopic 
model proposed by Bauge and co-workers  \cite{bau98, bau01a}   the absorption potentials for incident neutrons and protons respectively are plotted  in the energy range $10-100$ MeV for   \textrm{Pb}$^{208}$ . 

Circles indicate the original potential,
solid lines the
the optimum fit in the range $0 < r < 2 R_0$ for the proposed fractional derivative based potential (\ref{fcWS}).   
} }
\end{center}
\end{figure}
%-----------------------------------------------------------------------

For the simplified case of an  attractive contact interaction in spherical coordinates
\begin{eqnarray}
\label{contactW}
w_\delta &=& w(r, r') = -\frac{1}{4 \pi}\delta(|r-r'|)/\sqrt g 
\end{eqnarray}
the corresponding fractional potential $W_{WS}^\alpha$ based on the Woods-Saxon type density 
$\rho^{(\alpha)}_{WS}(r)$ 
 follows as:
\begin{eqnarray}
\label{fcWS}
W_{WS}^{\alpha}(r,\alpha) &=& W_\alpha a_0^\alpha \rho^{(\alpha)}_{WS}(r)\\
\label{denalpha2}
 &=& W_\alpha  \rho_0  \,  \textrm{Li}_{-\alpha}(-e^{(R_0-r)/a_0}) \qquad    0\leq \alpha \leq 1
\end{eqnarray} 
where $\alpha = \alpha(E)$ is a function of energy. 

In the top row of  figure  \ref{wfig2} we show a least square fit of the derived fractional Woods-Saxon potential  from (\ref{denalpha2}) with the absorption potential based on parameters  from Becetti and Greensleves  (\ref{parmsBec69})  for incident neutrons and protons respectively  in the energy range $10 \leq E \leq 50$ [MeV] for Pb$^{208}$.

The graphs of the adjusted parameters $R_0, a_0, W_{\alpha}, \alpha$, of the new fractional potential 
are  shown with solid lines in figures \ref{BecFitN} and  \ref{BecFitP}  for neutrons and protons respectively.  
The energy dependence
of $\alpha$ is nearly linear. Introducing a scaling factor $e_0$[MeV] we obtain:
\begin{eqnarray}
\label{alphamat}
\alpha(E) &\sim& 1-E/e_0   \qquad 0 \leq  E/e_0 \leq 1
\end{eqnarray} 
 The gross features of both potential types are similar, indicating that the fractional approach leads to reasonable results. A significant difference shows up in the intermediate energy region inside the nucleus.
Due to the simple form of of the classical  absorption potential, which is only a superposition  of the Woods-Saxon potential and its derivative, inside the nucleus there is only a constant contribution,  while the fractional 
analogue shows a definite dominant  decline with $r$ in the inner region.   

This behaviour is a direct consequence of the fractional approach which introduces a new quality named nonlocality when performing the convolution integral with the weakly singular kernel $w_L$ and thus
performing a infinite weighted sum of density values along the path.  

The fractional approach anticipates the development 
of more sophisticated microscopic optical model potentials, where 
an effective nucleon-nucleon interaction is folded with the matter density distribution \cite{kon03}.
Both mechanisms introduce nonlocal aspects.  

In the next section, we consider the consequences of a nonlocal approach.  
%--------------- figure becP
\begin{figure}[t]
\begin{flushright}
\includegraphics[width=\textwidth]{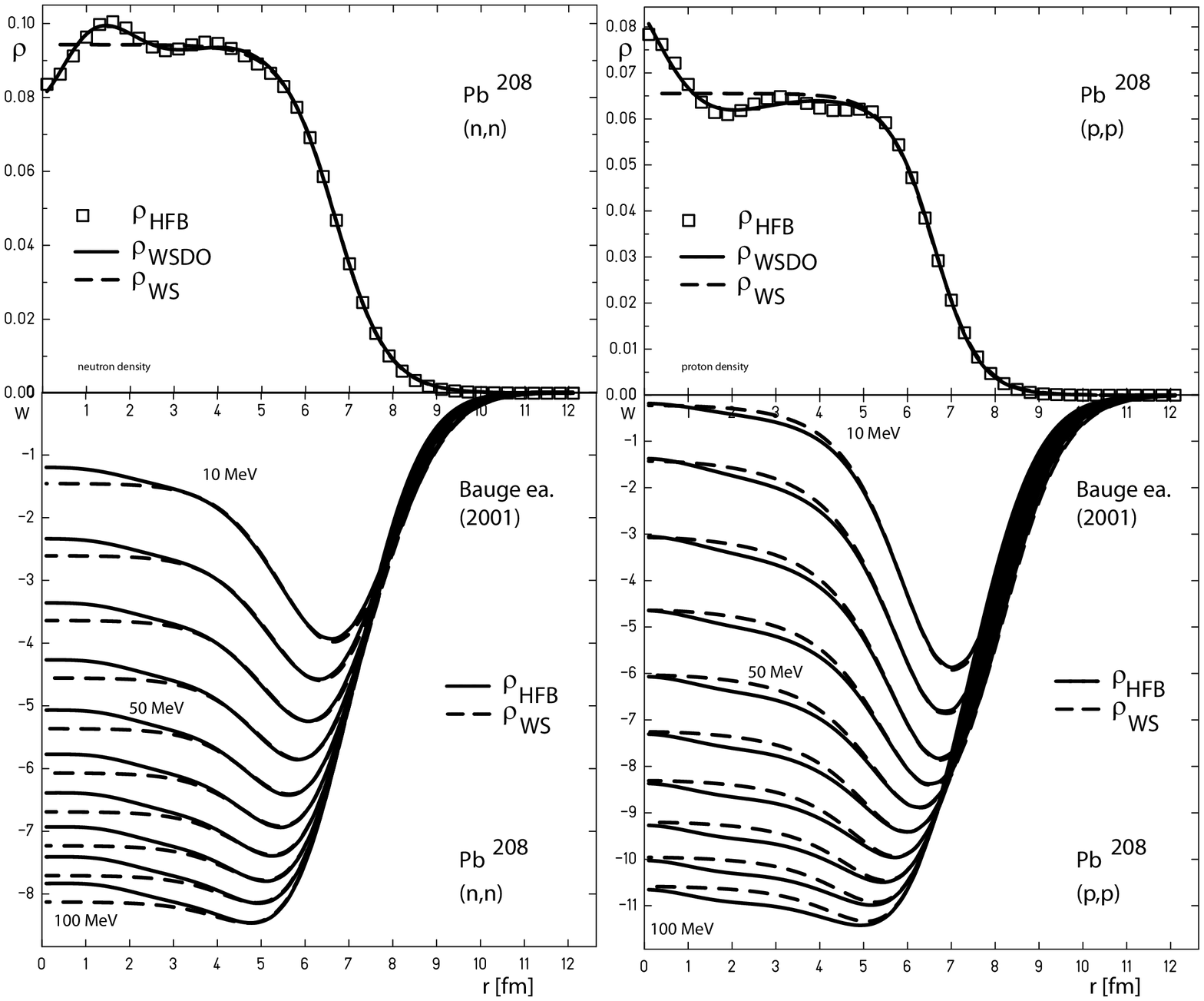}\\
\caption{
\label{BdensitiesFit}
{
Top: Fit result for densities $\rho_{HFB}$ of $Pb^{208}$ published by Bauge and co-workers  \cite{bau01b} (marked by dots)  with classical ($\alpha=0$) Woods-Saxon density  (\ref{dderivfrac2}) 
   (dashed line ) and 
from (\ref{fdDO}) for classical ($\alpha=0$)  Woods-Saxon with damped oscillation  admixture (solid line) 
for left/right neutrons/protons.

Bottom: The fitted densities served then as a basis to calculate the absorption potentials  
for $Pb^{208}$ within the energy range of 10-100 MeV 
with software package MOM published by Bauge and co-workers  \cite{bau01b} 
the solid line shows the resulting potentials for the original density $\rho_{HFB}$ 
the dashed curves show the deviation for the fitted classical Woods-Saxon density ($\alpha=0$)
and therefore the influence of oscillatory admixtures to the nucleon density on the absorption potential. Note the slope change in the inner region of the potential is almost missing for  neutrons
for $\rho_{WS}$ but sets in for protons already for $\rho_{WS}$ indicating the nonlocal character of the Coulomb force treatment.  
} }
\end{flushright}
\end{figure}
%-----------------------------------------------------------------------
\section{Nonlocality - a comparison of the fractional approach to microscopic optical models}

%-----------------------------------------------------------------------

The optical model potential is both nonlocal and energy dependent. There are different levels of
nonlocality:

On the highest level we may introduce nonlocal potentials and extend the Schr\"odinger equation 
(\ref{ctemp_se}) to an integro-differential equation as e.g. proposed by Perey and Buck \cite{per62}:
\begin{eqnarray}
\label{ctemp_senL}
(- \frac{\hbar^2}{2 m}\Delta -   E ) \Psi(\vec{r},E)  &=& -  \int dr' U(\vec{r},E) w(|r-r'|) \Psi(\vec{r},E) 
\end{eqnarray}
with a normalized non-singular kernel $w(x)$, e.g. a gaussian with a nonlocality spread $\sigma > 0$:
\begin{eqnarray}
\label{nskernel}
w(x) &=& \frac{1}{\sigma^3 \pi \sqrt{\pi}}e^{-x^2/\sigma^2} 
\end{eqnarray}

On a medium level we may consider folding the local density $\rho(\vec{r})$ with an appropriately
choses weight of Coulomb or effective nucleon-nucleon interaction $w$ according to (\ref{denVol}) in order to obtain a nonlocal potential. This level of nonlocality is realized
for many microscopic of semi-microscopic optical models. \cite{ jon63, hod67, rap79, kon03, mue11, kon14}.

Finally in the previous section we introduced an additional lower level of nonlocality by  folding of the density $\rho(\vec{r})$ with a  weakly singular kernel $w_L$ to obtain an intermediate fractional derivative  $\rho^{(\alpha)}(\vec{r})$.

As a consequence,  the proposed fractional optical potential (\ref{Vderiva})  is obtained by  first folding the local density $\rho_{WS}$ (\ref{ctemp_fn}) of Woods-Saxon type with the nonlocal fractional derivative kernel $w_L$ to obtain a nonlocal fractional density  
$\rho^{(\alpha)}_{WS}$ (\ref{dderivfrac2}).

In a second step within for reasons of simpicity this nonlocal  $\rho^{(\alpha)}_{WS}$ is  folded with an the attractive, but for now local contact potential $w_\delta(r, r')$ from (\ref{contactW})  in spherical coordinates to obtain the fractional absorption potential $W^{\alpha}$. 
\begin{eqnarray}
\label{V000}
W^{\alpha}(r, \alpha) 
&\sim& (w_\delta  \ast  w_L  \ast \rho_{WS})(r) 
\qquad\qquad 0\leq \alpha \leq 1, \alpha \in \mathbb{R}
\end{eqnarray}
Of course, nonlocal potential kernels may be used too.
 
In this section we will discuss the influence of this new nonlocality on the density level and compare it with  the nonlocality on the potential level  proposed
within a microscopic model. As an example  in the following we compare with the  well established  microscopic model  proposed by Bauge and co-workers  \cite{bau98, bau01a}. 

In general within the microscopic optical models there are two main optimizations included.

First, attempts are made to generate a more sophisticated density distribution for the nucleons and 
second, an appropriate collective effective nucleon potential is chosen, which is used for the
folding procedure of the density to yield appropriate collective optical potential contributions. 
This folding procedure introduces  the new concept of nonlocality into the optical model, provided that  an appropriate nonlocal kernel is used.   

In order to generate the imaginary optical model potential contribution we use the
code MOM \cite{bau01b}, which comes with a test file with precalculated neutron/proton densities  for 
$Pb^{208}$ which are folded with a modified effective nucleon interaction which yields corresponding optical model potentials.

%--------------- figure becP
\begin{figure}[t]
\begin{flushright}
\includegraphics[width=\textwidth]{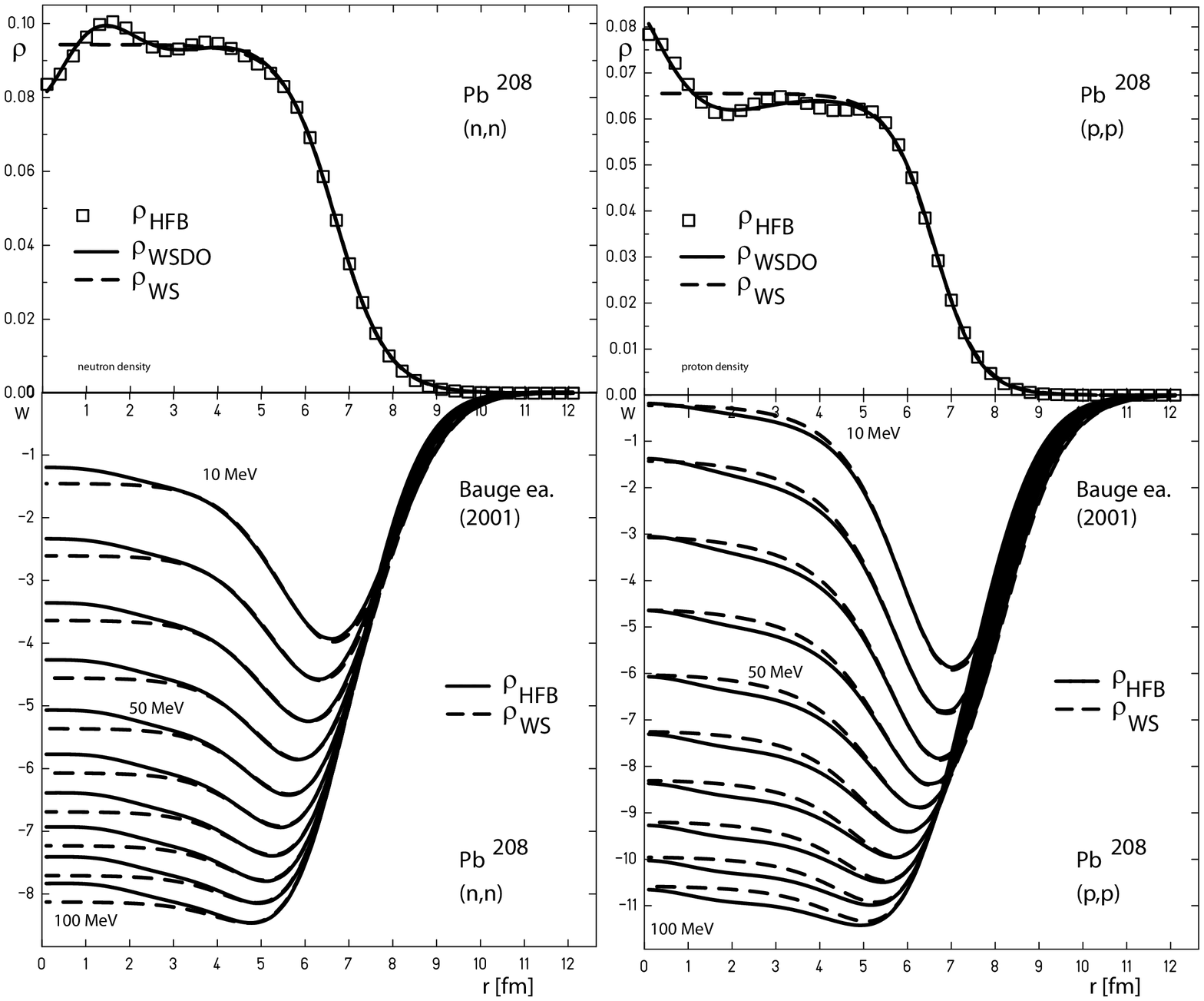}\\
\caption{
\label{BdensitiesFit}
{
Top: Fit result for densities $\rho_{HFB}$ of $Pb^{208}$ published by Bauge and co-workers  \cite{bau01b} (marked by dots)  with classical ($\alpha=0$) Woods-Saxon density  (\ref{dderivfrac2}) 
   (dashed line ) and 
from (\ref{fdDO}) for classical ($\alpha=0$)  Woods-Saxon with damped oscillation  admixture (solid line) 
for left/right neutrons/protons.

Bottom: The fitted densities served then as a basis to calculate the absorption potentials  
for $Pb^{208}$ within the energy range of 10-100 MeV 
with software package MOM published by Bauge and co-workers  \cite{bau01b} 
the solid line shows the resulting potentials for the original density $\rho_{HFB}$ 
the dashed curves show the deviation for the fitted classical Woods-Saxon density ($\alpha=0$)
and therefore the influence of oscillatory admixtures to the nucleon density on the absorption potential. Note the slope change in the inner region of the potential is almost missing for  neutrons
for $\rho_{WS}$ but sets in for protons already for $\rho_{WS}$ indicating the nonlocal character of the Coulomb force treatment.  
} }
\end{flushright}
\end{figure}
%-----------------------------------------------------------------------

In the bottom part of figure \ref{wfig2} we compare the absorption potentials from  Bauge and co-workers  \cite{bau01a} (circles) with the fractional Woods-Saxon type potential (\ref{Vderiva})
(solid lines). 
Both graphs agree much better over the whole energy range than in the case of
Becetti's and Greensleves' potential see upper part if figure \ref{wfig2}. 

For both cases we observe a 
similar slope behaviour inside the target nucleus. Within the classical approach this 
behaviour sets in when the simple superposition of local volume and surface attribution is 
replaced by a more sophisticated non-local treatment of the potential in the microscopic models.   

In the fractional approach nonlocality enters right from the beginning as a characteristic feature of the fractional derivative definition itself.
On the other hand, in the classical derivation of the absorption potential according to microscopic models, nonlocality enters 
in  MOM from Bauge and co-workers  \cite{bau01a}
via the use of a nonlocal kernel of e.g. Gogny-type \cite{dec80} folding with a nucleon density, derived by HFB-calculations.

Formally written as 
\begin{equation}
\label{Vgogny}
W(r, \alpha) \sim ( w_{\textrm{Gogny}}  \ast \rho_{HFB})(r)
\end{equation}
where $\rho_{HFB}(r)$ is the input density .

Of course the question remains, what is the influence of the input density for the resulting
absorption potential within the microscopic model considered.

In the upper row  of  figure \ref{BdensitiesFit} $\rho_{HFB}$ 
from MOM from Bauge and co-workers  \cite{bau01a}
is plotted with squares. We observe dampled oscillatory admixtures to the simple Woods-Saxon
density $\rho_{WS}$ (dashed lines). 

%--------------- figure bau2
\begin{figure}
\begin{center}
\includegraphics[width=\textwidth]{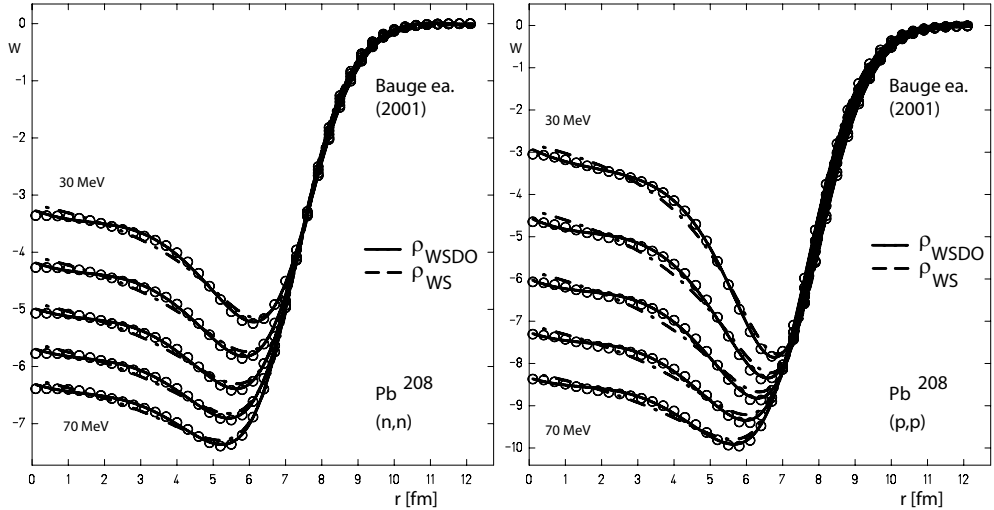}\\
\caption{
\label{DDDFit}
{
Fit result of the energy dependent absorption potential (circles) from Bauge and co-workers  \cite{bau01a} 
for $Pb^{208}$ within the intermediate energy range 30-70 MeV 
with the fractional  Woods-Saxon potential  $W^\alpha_{WS}$ (\ref{denalpha2}) (dashed lines),  and with  the fractional Woods-Saxon plus damped oscillation potential  $W^\alpha_{WSDO}$ (\ref{fvDO})(solid lines). The corresponding error is sketched in figure \ref{DDDFitNPERR}.
 }}
\end{center}
\end{figure}
%-----------------------------------------------------------------------

In the bottom row in figure \ref{wfig2} we compare the influence of oscillatory admixtures 
of the input density for the microscopic model calculations. The resulting absorption potential 
is plotted, which was calculated with MOM  for the input densities of the standard Woods-Saxon type $\rho_{WS}$ and  $\rho_{HFB}(r)$.  

The slope of the calculated absorption potential inside the  
nucleus is a direct property of the use of a microscopic model. Exactly this feature is missing using
Beccetti's  superposition ansatz, while the fractional potential $W^\alpha$ already 
shows this behaviour. 

The presented results show, that the fractional approach
 implements right from the scratch while 
performing the step from $\rho_{WS}$ to $\rho^{(\alpha)}_{WS}$ 
an  nonlocal aspect to the absorption potential behaviour, which in the classical approach has to be modeled introducing additional nonlocal methods, e.g. folding with effective nucleon-nucleon interactions.  The fractional Woods-Saxon potential, which is based on the nonlocal
fractional Woods-Saxon type density,   already anticipates nonlocal
effects, which later in the development of the optical model occurred using more sophisticated microscopic models.  
 
\section{An extended Woods-Saxon potential with a damped oscillation term}

Motivated by the functional behaviour of the more sophisticated density distribution from HFB-calculations,
we now present a reasonable extension of the pure Woods-Saxon type densities, which will yield 
an even better agreement with the potentials presented by Bauge and co-workers  \cite{bau01a}.

%--------------- figure becNPERR
\begin{figure}[t]
\begin{center}
\includegraphics[width=\textwidth]{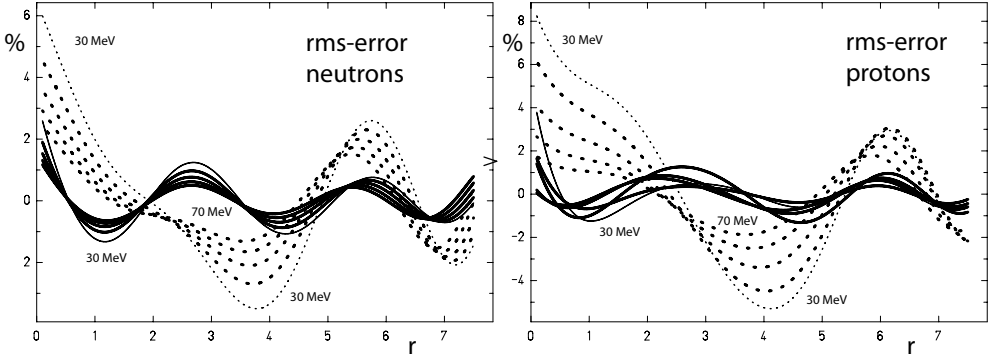}\\
\caption{
\label{DDDFitNPERR}
{
Root mean square error for the fits plotted  in figure  \ref{DDDFit}. Dashed lines show the error for the 
standard Woods-Saxon potential (\ref{denalpha2}), solid lines show the error for for the extended Woods-Saxon plus damped oscillation potential (\ref{fvDO}). The overall agreement with the potential increases by a factor 2-3.
 }}
\end{center}
\end{figure}
%-----------------------------------------------------------------------

Extending the Woods-Saxon density  (\ref{ctemp_fn}) by a damped oscillatory part via
\begin{eqnarray}
\label{WSDO}
\rho_{WSDO} &=& \rho_{WS} + \rho_{DO} 
\end{eqnarray}
where $\rho_{DO}$ is given by
\begin{eqnarray}
\label{DO}
\rho_{DO}(r) &=& \rho_{0} \, e^{-k r} \cos(m r + \phi)   \qquad \quad  \{k,m,\phi  \} \in \mathbb{R}, \, \alpha, k > 0 
\end{eqnarray}
with a damping factor $k>0$, oscillator frequency m and a phase $\phi$.

Since the fractional derivative of the exponential is easily calculated with the help of (\ref{dintfrac}):
\begin{eqnarray}
\label{dexp0}
\frac{\partial^\alpha}{\partial r^\alpha} e^{-k r + \phi}&=&  
k^\alpha
e^{-k r + \phi}  \qquad \qquad \quad  \{k , \phi  \} \in \mathbb{C}, \, \Re(\alpha, k) > 0 
\end{eqnarray}
we obtain for the fractional derivative of $\rho_{DO}$:
\begin{eqnarray}
\label{fdDO}
\rho^{(\alpha)}_{DO}(r) &=&  \frac{1}{2} \rho_{0} \,  e^{-k x} \left(   e^{i (m x+\phi)} (k-i m)^\alpha + e^{-i (m x+\phi)} (k+i m)^\alpha \right) \nonumber \\
 && \qquad \qquad \qquad \qquad\qquad  \{k,m,\phi  \} \in \mathbb{R}, \, \alpha, k > 0 
\end{eqnarray}
which is a real quantity, since for the complex conjugate $\overline{ \rho ^{(\alpha)}_{DO}  }(r) = \rho^{(\alpha)}_{DO}(r)$ holds. 

For the local contact kernel $w_{\delta}$  the fractional Woods-Saxon plus 
damped oscillation potential follows as:
\begin{eqnarray}
\label{fvDOnull}
W_{WSDO}^{\alpha}&(r, \alpha)& = 
W_\alpha a_0^\alpha  (w_\delta  \ast  \rho ^{(\alpha)}_{WSDO} )(r) \\
&=& a_0^\alpha  \Biggl( W_\alpha    \textrm{Li}_{-\alpha}(-e^{(R_0-r)/a_0})  +  \nonumber \\
& + &
\frac{W_\sim}{2} e^{-k x} \left(   e^{i (m x+\phi)} (k-i m)^\alpha + e^{-i (m x+\phi)} (k+i m)^\alpha \right) \biggr)  \nonumber \\
\label{fvDO}
&=&  a_0^\alpha  \Biggl( W_\alpha  \textrm{F}_{-\alpha-1}((R_0-r)/a_0)   + \nonumber \\
& + &
\frac{W_\sim}{2} e^{-k x} \left(   e^{i (m x+\phi)} (k-i m)^\alpha + e^{-i (m x+\phi)} (k+i m)^\alpha \right) \biggr) \nonumber \\
&&
\qquad\qquad\qquad\qquad,\qquad 0\leq \alpha \leq 1, \alpha \in \mathbb{R}
\end{eqnarray}

%---------------------- WS + exp
\begin{table}
\scalebox{1}{
{\begin{tabular}{l|rrrrrrrrr}
\hline\noalign{\smallskip}
$W_{index}^{\alpha}(r)$ &$\alpha$&  $R_0$ & $a_0$ & $W_\alpha$& $W_\sim$& $k$ & $m$ &$\phi$&error\\
\hline\noalign{\smallskip}
 neutrons  & & & & & & &&&  \\
WS  &0.253 &7.35&0.670&10.08&           &      &   &   &7.2E-2 \\
WSDO  &0.285 &7.21&0.744&10.44&0.22&0.076&0.96&&2.7E-2 \\
WSDO(P) &0.284  &7.22&0.744&10.42&0.22&0.075&0.955&0.036& 2.7E-2 \\
protons   &      &      &          &         &          &        &       && \\
WS       & 0.380 &7.69&0.680&18.16&         &        &     &  &1.4E-1 \\
WSDO   &0.416  &7.54&0.732&23.62&3.22&-0.18&0.41&& 2.7E-2 \\
WSDO(P) &0.420  &7.54&0.746&24.16&2.98&-0.13&0.42& -0.24& 2.4E-2 \\
\hline\noalign{\smallskip}
\end{tabular} }}
\label{ws222}
\caption{Example fit parameters for the absorption potential   figure  \ref{DDDFit} for 40 MeV projectile
energy. Listed are values for $W_{WS}^{\alpha}(r)$, the  
fractional standard Woods-Saxon potential $W_{WS}^{\alpha}(r)$   (\ref{denalpha2}), $W_{WSDO}^{\alpha}(r)$ and $W_{WSDO(P)}^{\alpha}(r)$, the extended Woods-Saxon plus damped oscillation potential (\ref{fvDO}) without/with phase shift. }
\end{table} 
%----------------------
In figure  \ref{DDDFit} we compare the fractional standard Woods Saxon potential (\ref{fcWS}) and the extended Woods-Saxon potential   (\ref{fdDO}) in the region 10-100 MeV projectile energy for the 
double magic $Pb^{208}$
with the semi-microscopic absorption potential from \cite{bau01b}. In figure  \ref{DDDFitNPERR} 
we plot the error for 30-70 MeV and in table 1 we list the fit parameters for the case
40 MeV  in order to give an impression of the parameter change when applying the different model potentials.

Taking into account possible
density fluctuations by extending the fractional Woods-Saxon potential by an 
additional term for fractional damped oscillations 
results in a potential model, which is flexible enough, to reduce   
the difference between the extended fractional and microscopic approach by an 
additional factor 2-3.  

This allows a variation of the resulting fractional absorption potential similar to semi-microscopic  models and opens a wide range of possible applications.

\section{The fractional global parameter set}
%--------------- figure becN
\begin{figure}[t]
\begin{center}
\includegraphics[width=\textwidth]{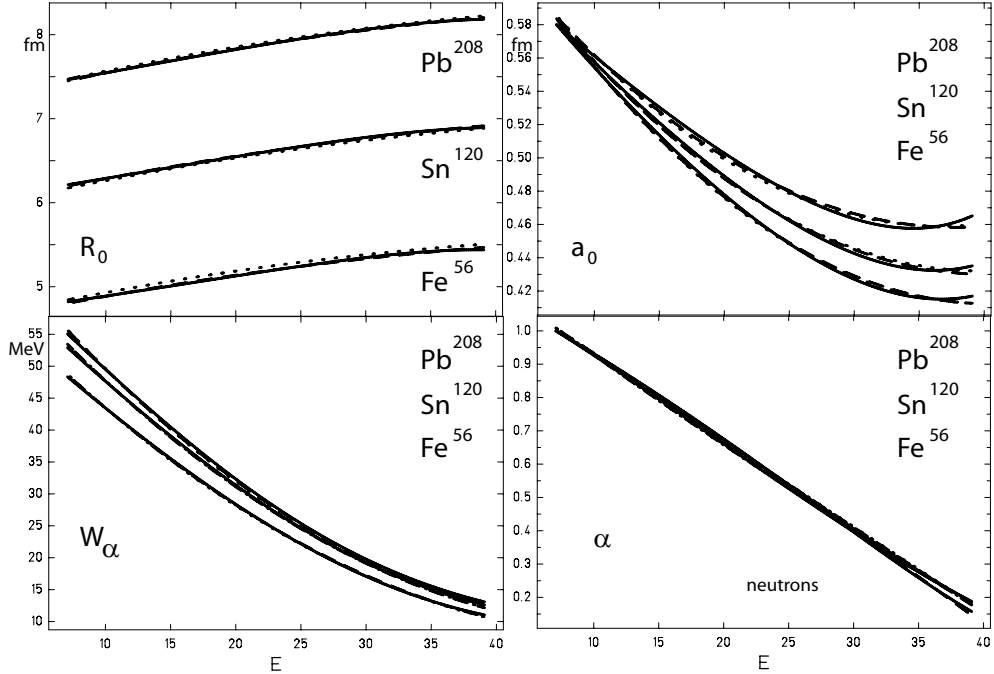}\\
\caption{
\label{BecFitN}
{
For $R_0$, $a_0$, $W_{\alpha}$ and $\alpha$ the global fit of the absorption potential according to (\ref{omp_fn}) (solid line)  with 
the potential based on the parameter set proposed by Becetti and Greensleves according to (\ref{parmsBec69}) (dashed line) is plotted for 3 different nuclei (Pb, Sn, Fe) in the energy range $10 \leq E \leq 40$  
for incident  neutron projectiles.} }
\end{center}
\end{figure}
%-----------------------------------------------------------------------
%--------------- figure becP
\begin{figure}[t]
\begin{flushright}
\includegraphics[width=\textwidth]{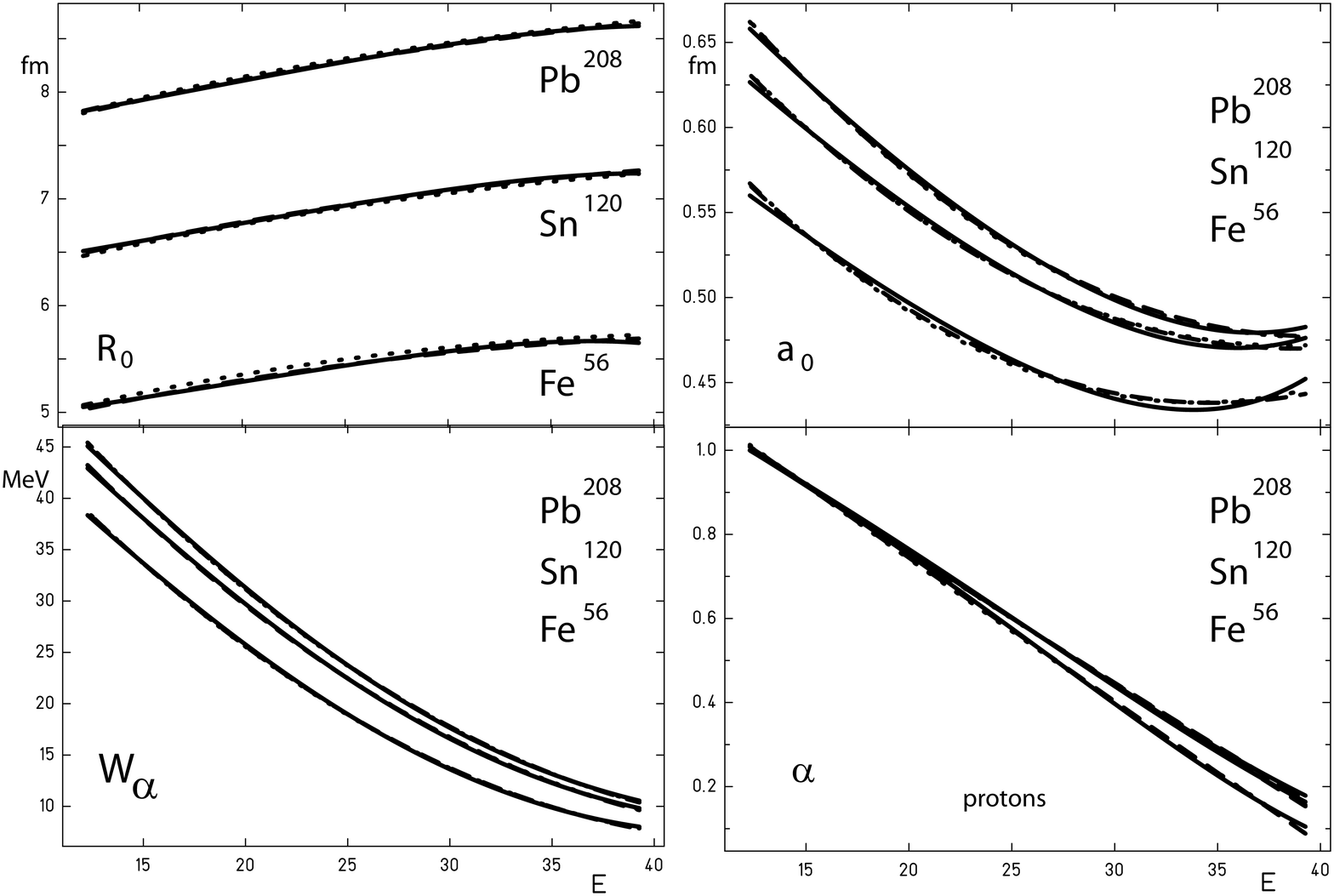}\\
\caption{
\label{BecFitP}
{
For $R_0$, $a_0$, $W_{\alpha}$ and $\alpha$ the global fit of the absorption potential according to (\ref{omp_fn}) (solid line)  with 
the potential based on the parameter set proposed by Becetti and Greensleves according to (\ref{parmsBec69}) (dashed line) is plotted for 3 different nuclei (Pb, Sn, Fe) in the energy range $10 \leq E \leq 40$  
for incident proton projectiles.} }
\end{flushright}
\end{figure}
%-----------------------------------------------------------------------

%---------------------- neutrons 32
\begin{table}
\scalebox{0.9}{
{\begin{tabular}{r|rrrr}
\hline\noalign{\smallskip}
neutrons &  & $\mu $    &  & \\
\hline\noalign{\smallskip}
$b_{ijk}$ &  $\alpha$ & $R_0$& $a_0$ & $W_{\alpha}$\\
\noalign{\smallskip}\hline\noalign{\smallskip}
$b_{000}$&1.16002&3.23855&6.31512E-1&5.83134E1\\
$b_{100}$&-2.09028E-2&2.79605E-2&-7.64432E-3&-1.97058\\ 
$b_{200}$&-1.39729E-4&-2.26493E-4&1.09326E-4&1.87064E-2\\ 
$b_{010}$&-5.36764E-2&-4.36717E-1&2.12094E-1&4.68971E1\\ 
$b_{110}$&1.27116E-2&2.65314E-2&-1.84521E-2&2.50232E-1\\ 
$b_{210}$&1.88482E-4&-2.30697E-5&1.4966E-4&-2.36055E-2\\ 
$b_{020}$&-1.05737&5.21175&-2.16473&-2.15378E1\\
$b_{120}$&1.20967E-1&-6.0039E-1&2.65391E-1&3.2181\\ 
$b_{220}$&-3.00816E-3&1.2824E-2&-6.02127E-3&-6.33512E-2\\ 
$b_{001}$&9.60025E-4&2.61872E-2&4.3146E-4&2.39938E-2\\ 
$b_{101}$&-1.41237E-4&2.19149E-4&-7.05559E-5&-4.22499E-3\\ 
$b_{201}$&2.5512E-6&-3.698E-6&1.20596E-6&8.4832E-5\\
$b_{011}$&-2.30815E-3&6.80988E-3&-2.49754E-4&9.36127E-2\\ 
$b_{111}$&3.34947E-4&-9.15282E-4&1.39433E-4&-5.23342E-3\\ 
$b_{211}$&-7.41202E-6&1.77282E-5&-3.16304E-6&5.68913E-5\\ 
$b_{021}$&1.36172E-2&-5.81241E-2&1.76318E-2&-1.02795E-1\\ 
$b_{121}$&-1.76209E-3&7.27927E-3&-2.49776E-3&-1.11364E-2\\ 
$b_{221}$&4.02172E-5&-1.61211E-4&5.92301E-5&3.87674E-4\\ 
$b_{002}$&-1.70111E-6&-3.67824E-5&-7.23149E-7&-3.58834E-5\\ 
$b_{102}$&2.47862E-7&-4.0281E-7&1.26007E-7&6.93429E-6\\ 
$b_{202}$&-4.31213E-9&6.61254E-9&-2.1266E-9&-1.36821E-7\\ 
$b_{012}$&4.73126E-6&-1.62584E-5&-9.71364E-7&-3.56636E-4\\ 
$b_{112}$&-7.41566E-7&2.40676E-6&-3.18572E-7&2.02272E-5\\ 
$b_{212}$&1.60083E-8&-4.79778E-8&8.58725E-9&-2.79077E-7\\ 
$b_{022}$&-2.99806E-5&1.31688E-4&-2.86779E-5&9.99785E-4\\ 
$b_{122}$&4.14497E-6&-1.7493E-5&5.08152E-6&-2.76855E-5\\ 
$b_{222}$&-9.54636E-8&3.92787E-7&-1.27997E-7&2.01557E-8\\
\noalign{\smallskip}\hline\noalign{\smallskip}
\end{tabular} }}
\label{tab:ompneu}
\caption{Fit parameters $b_{ijk}$ according to (\ref{omp_fn}) for neutrons}
\end{table} 
%----------------------

%---------------------- protons 32
\begin{table}
\scalebox{0.9}{
{\begin{tabular}{r|rrrr}
\hline\noalign{\smallskip}
 protons & &  $\mu$&  & \\
\hline\noalign{\smallskip}
$b_{ijk}$ & $\alpha$ & $R_0$& $a_0$ & $W_{\alpha}$\\
\noalign{\smallskip}\hline\noalign{\smallskip}
$b_{000}$&1.45738&2.97928&6.91304E-1&6.22561E1\\
$b_{100}$&-3.57272E-2&5.61629E-2&-1.73707E-2&-2.54804\\
$b_{200}$&-1.52847E-5&-6.78655E-4&2.71946E-4&2.87836E-2\\
$b_{010}$&-1.73192&2.17068&3.82327E-2&1.89994E1\\
$b_{110}$&1.55702E-1&-2.18343E-1&6.80405E-2&3.28564\\
$b_{210}$&-2.0218E-3&4.64791E-3&-1.47557E-3&-8.31024E-2\\
$b_{020}$&2.79229&-3.24036&9.68804E-1&2.80201E1\\
$b_{120}$&-2.6603E-1&3.36065E-1&-1.01722E-1&-3.36481\\
$b_{220}$&4.72882E-3&-7.92969E-3&2.43325E-3&9.59847E-2\\
$b_{001}$&2.60798E-3&2.47759E-2&1.33373E-3&5.46996E-2\\
$b_{101}$&-2.64892E-4&4.22512E-4&-1.35713E-4&-5.91297E-3\\
$b_{201}$&4.94074E-6&-7.91767E-6&2.52546E-6&1.15096E-4\\
$b_{011}$&-6.62564E-3&1.46379E-2&-3.2664E-3&3.00421E-2\\
$b_{111}$&7.43875E-4&-1.52236E-3&3.51202E-4&-6.14844E-4\\
$b_{211}$&-1.81707E-5&3.72658E-5&-9.11869E-6&-7.04581E-5\\
$b_{021}$&6.71699E-3&-2.75892E-2&5.68185E-3&-1.28594E-1\\
$b_{121}$&-8.09413E-4&2.78814E-3&-5.82523E-4&1.03988E-2\\
$b_{221}$&2.20862E-5&-6.63321E-5&1.45556E-5&-1.46995E-4\\
$b_{002}$&-6.07232E-6&-2.85172E-5&-4.5608E-6&-1.24689E-4\\
$b_{102}$&6.01584E-7&-1.25619E-6&4.47185E-7&1.29714E-5\\
$b_{202}$&-1.14806E-8&2.48641E-8&-8.89248E-9&-2.55434E-7\\
$b_{012}$&3.11991E-5&-9.71514E-5&3.37022E-5&3.20899E-4\\
$b_{112}$&-3.159E-6&9.46288E-6&-3.2767E-6&-3.44315E-5\\
$b_{212}$&7.16834E-8&-2.12167E-7&7.35789E-8&8.89428E-7\\
$b_{022}$&-6.24725E-5&2.38078E-4&-8.37765E-5&-7.11391E-4\\
$b_{122}$&6.24831E-6&-2.2947E-5&8.04581E-6&7.14510E-5\\
$b_{222}$&-1.43356E-7&5.07704E-7&-1.78026E-7&-1.69351E-6\\
\noalign{\smallskip}\hline\noalign{\smallskip}
\end{tabular} }}
\label{tab:omppro}
\caption{Fit parameters $b_{ijk}$ according to  (\ref{omp_fn})  for protons}
\end{table} 
%----------------------

In the previous section we have applied the generated fractional model potential to a single nucleus using data for lead $Pb^{208}$ as an example. Now we want to extend the model parameter set to a wider range of nuclear targets.  

Optimized parameter sets for the classical models have been reported for nucleon elastic scattering e.g. by
\cite{bec69, rap79, wal86, var91, kon03} for a large variety of nuclei and energies. These were obtained by a fit with experimental cross sections and lead to corresponding optical potentials.

We will use a direct approach by fitting parameters for the previously derived fractional optical model potential directly with the classical optimal potential parameters  given by Becetti and Greensleves according to (\ref{parmsBec69}).  

We perform a two-step procedure: First for every valid projectile, energy and nuclear asymmetry combination we  fit 
the classical absorption potential 
with the fractional
Woods Saxon model. In the second step the resulting multi-dimensional point cluster is then  fitted with an appropriately chosen fitting function.

As a fitting function ansatz which will minimize the error of the given 
fractional  absorption potential  (\ref{denalpha2}) cluster we use an quadratic ansatz with parameters nucleon number $A$, asymmetry $I = (N-Z)/A$ and energy E:
\begin{eqnarray}
\label{omp_fn}
f^{\mu}_{\tau}(A, I, E) = \sum_{i,j,k=0}^{i,j,k=2} b_{ijk}E^i I^j A^k, &&\\ 
&& \mu \in \{\alpha,a_0,R_0, W_{\alpha}\} \nonumber\\
&& \tau \in \{\textrm{neutrons, protons}\} \nonumber
\end{eqnarray} 
In tables 2 and 3 we have listed the adjusted $b_{ijk}$  
for $\alpha$,  $R_0$, $a_0$ and $W_{\alpha}$ for  neutrons and protons respectively. In figures \ref{BecFitN} and
\ref{BecFitP} the corresponding graphs for the optimum parameter sets are plotted. 
The corrspondence of our fractional model curves with the values proposed by Becetti
(\ref{parmsBec69})
 is remarkable.

For parameters $\alpha, R_0, W_{\alpha}$ of the fractional model  we obtain an almost linear behaviour throughout the periodic table in the proposed energy region. It is noteworthy that especially the fractional parameter $\alpha$ shows a dominant  linear dependence from energy almost
independently of the target nucleus.

Of course this is only a coarse adjustment of the fractional parameters, because we compared results only on the potential level. 

A next important step for future development will be a fine tuning of the fractional parameters based on  the use of experimentally measured cross sections.

\section{A-posteriori legitimation of the classical approach}
Despite the fact, that the fractional derivative is the correct method to realize  the intended smooth transition from surface to volume absorption potentials there remain open questions: 

Why does the classical description of  the same phenomenon  
in terms of a simple superposition of the first and zeroth derivative lead to comparable good results? 

Is this a special case for functions of Woods-Saxon type only? 

In the following we will give an answer presenting a different interpretation of a fractional derivative, 
which is based on an infinite series expansion of the fractional derivative in terms of integer derivatives. 

At first glance it is tempting to assume the fractional calculus approach to derive a reasonable
optical potential as a fractional derivative of the nuclear density function would be  a simple series expansion
of the same derivative in terms of integer derivatives 
\begin{equation}
\rho^{(\alpha)}(x)  \stackrel{?}{=}  \sum_{i=0}^\infty \tilde{c}_i(\alpha) \rho^{(i)}(x)
\end{equation}
with spatially independent coefficients $\tilde{c}_i(\alpha)$.
The classical approach was then interpreted as a truncation of this series to two terms only, namely $i\ leq 1$ for the zeroth and first derivative of the density 
function. 

In fractional calculus things are not that simple. 
 One of the premises for any reasonable definition of a  fractional derivative is that  the 
 fractional extension of the classical  Leibniz product rule is to be fulfilled, which is given by:
\begin{equation}
\label{leibniz}
 (\psi \, \chi )^{(\alpha)}(x) = \sum_{j=0}^\infty    
 % ssssss s t a r t binomial
\left(\begin{array}{c}
\alpha \\ j \\
\end{array} \right)
 % ssssss e n d  binomial
       \psi^{(\alpha-j)}(x)  \chi^{(j)}(x) 
\end{equation}

Rewriting  the analytic function $\rho(x)$ as a general product:
\begin{equation}
\rho(x) = \lim_{\beta \rightarrow  0}  x^{\beta} \rho(x) = x^0 \rho(x) \qquad \beta, x \geq 0
\end{equation}
or equivalently setting $\psi(x) = x^0 = 1$ and consequently interpreting the term $\partial_x^{\alpha-j}  \psi(x)$ as the fractional integral of a constant function proves its  $x$ dependence  even for the  case of integer $\alpha$, $\alpha = n \in \mathbb{N}$.  
So, although that the fractional extension of the Leibniz product rule is the correct starting point for a series expansion of the fractional derivative in term of integer derivatives, we obtain 
space dependent  coefficients $c_i(\alpha, x)$. 
\begin{equation}
\label{trueLeib}
\rho^{(\alpha)}(x)  \stackrel{!}{=}  \sum_{i=0}^\infty c(\alpha, x) \rho^{(i)}(x)
\end{equation}

Nevertheless, we propose two approaches to eliminate  the spatial dependence of the coefficients in (\ref{trueLeib}). 
The first approach is based on the Gaussian least squares method determining the coefficients
 $\tilde{c}_i(\alpha)$ within
a given interval $[a,b]$ as solutions of the overdetermined system of equations ($N \gg M \in \mathbb{N}$):
\begin{eqnarray}
\label{qqgauss}
\delta \sum_{i=0}^{(b-a)/N}  \left( \rho^{(\alpha)}(x_i) - \sum_{i=0}^M \tilde{c}_i(\alpha) \rho^{(i)}(x)\, |_{x_i}  \right)^2    &=& 0
\end{eqnarray}
Especially for  the Woods-Saxon type functions we will focus on the vicinity of $x = x_s = R_0 $, since only at this point we expect a significant contribution of higher order integer derivatives, while for $x \rightarrow 0$ and $x \rightarrow \infty$ the higher order derivatives ($i>0$) are negligible.  We therefore postulate, that a valid comparison of contributions of higher order derivatives makes sense
only in the vicinity of  $R_0$.

\begin{figure}[t]
\begin{center}
\includegraphics[width=\textwidth]{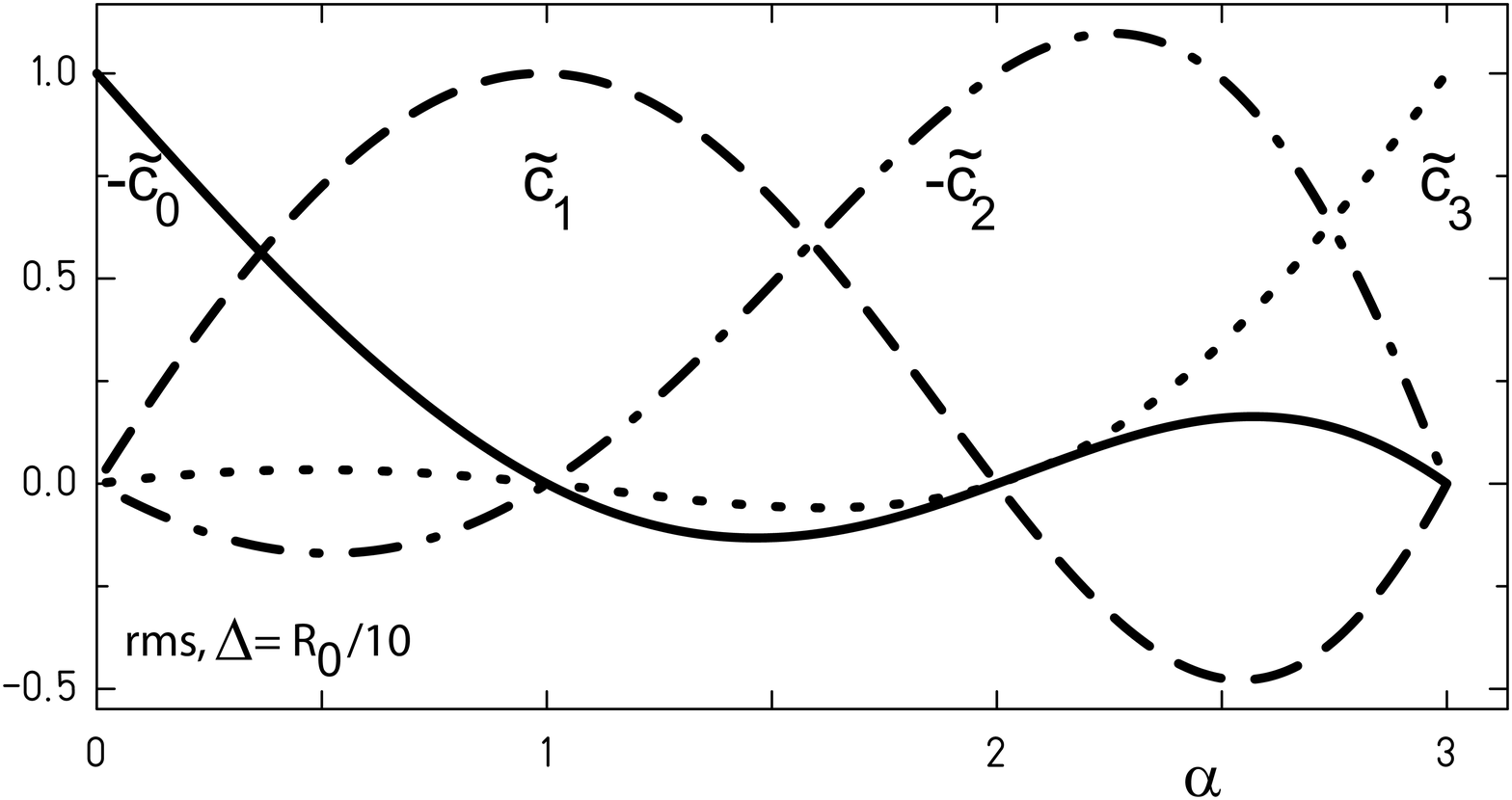}\\
\caption{
\label{wfigav}
{
For $^{208}\textrm{Pb}$ coefficients $\tilde{c}_j$ as a function of $\alpha$ from optimum fit of the fractional Woods-Saxon potential $ \rho^{(\alpha)}_{WS}(r)$ from (\ref{dderivfrac2}) in the vicinity $\Delta= R_0/10$  of $R_0$. Signs are adjusted such that $\tilde{c}_j(\alpha)=1$ for $\alpha=j$. 
The relevant region $0 < \alpha < 1$ may be directly compared with the classical linear ansatz e.g. (\ref{linearAlpha})  
or (\ref{parmsBec69}) 
} }
\end{center}
\end{figure}

Setting $M=4$, $a = R_0 - \Delta$, $b = R_0 + \Delta$ with  $\Delta = 0.1 R_0$ and $N=100$ we obtain
the set $\{\tilde{c}_0(\alpha) ,\tilde{c}_1(\alpha) ,\tilde{c}_2(\alpha) ,\tilde{c}_3(\alpha) \}$. In figure
\ref{wfigav} the parameter values for $Pb^{208}$ are plotted.

Before we discuss this result, let us take a look at  the second approach:

We will derive an alternative definition of a fractional derivative in terms of a sum of integer derivatives localized at a given position $x = x_s$, which allows a comparison with the above presented first approach. 
\begin{eqnarray}
\label{caa3}
\frac{ \partial^\alpha}{ \partial x^\alpha} f(x) |_{x=xs}&=& \sum_{j=0}^{\infty} c_j(x) \frac{\partial^j}{ \partial x^j} f(x) \quad |_{x=xs}
 \qquad    0\leq \alpha \leq 1\\
&=&c_0(x) f(x) + c_1(x) \frac{\partial}{ \partial x} f(x) + I(x,\alpha)\quad |_{x=xs}  
\end{eqnarray}
with a residual  $I(x_s,\alpha)$.

%--------------- figure frc
%-----------------------------------------------------------------------

We will then demonstrate, that the hitherto used classical approach covers the first two terms of this series expansion only, which at first seems quite a poor approximation. After that we will then derive an error estimate and will deduce, that the contribution of higher order terms in the series expansion are surprisingly small for
Woods-Saxon type functions for $\alpha$ in the range  $0 \leq \alpha \leq 1$.\\ 
For  Woods-Saxon type densities there are two conditions fulfilled:\\
First there exists a mirror point $x_s$ with the property:
\begin{eqnarray}
\label{qfirstcc}
2 \rho(x_s) &=& \rho(x_s + h) + \rho(x_s-h) 
\end{eqnarray}
In the special case of the  Woods-Saxon type density $x_s = R_0$.

Second, we have an asymptotic development of the form  $\lim_{r \rightarrow \infty} \rho_{WS} = 0$  and
 introducing the residual  $R(\alpha, x_s)$
\begin{eqnarray}
\label{dddsecond}
R(\alpha, x_s) &=& \partial_{x_s} \int_{x_s}^\infty \!\!\!\!dh \frac{1}{h^\alpha}\rho(x+h)  \, < \epsilon
\end{eqnarray}
we have  $R(\alpha, x_s)  < \epsilon $  such that
\begin{eqnarray}
\partial_x \int_0^\infty \!\!\!\!dh \frac{1}{h^\alpha}\rho(x+h) &=& 
\partial_x \int_0^{x_s} \!\!\!\!dh \frac{1}{h^\alpha}\rho(x+h) + R(\alpha, x_s) \\
\label{qq22}
&\approx& 
\partial_x \int_0^{x_s} \!\!\!\!dh \frac{1}{h^\alpha}\rho(x+h) 
\end{eqnarray}
That indeed  shows
that  for the  Woods-Saxon density $\rho_{\textrm{WS}}$ the residual is nothing else but the upper incomplete polylogarithm and we can give an upper estimate for $\epsilon $ using properties of  the exponential integral  $E_\alpha (x)$
\begin{eqnarray}
R(\alpha, x_s) &=&  \partial_{x_s} \frac{1}{\Gamma(1-\alpha)} \int_{x_s}^\infty \!\!\!\!dh \frac{1}{h^\alpha}\frac{1}{1 + e^{(R_0-x)/a_0}} \\
&<&  \partial_{x_s} \frac{1}{\Gamma(1-\alpha)} \int_{x_s}^\infty \!\!\!\!dh \frac{1}{h^\alpha}e^{-(R_0-x)/a_0} \\
&=& e^{\frac{R_0-x_s}{a_0}}\frac{x_s^{-\alpha}}{\Gamma(1-\alpha)} (1 + \frac{x_s}{a_0}e^{x_s/a_0}E_\alpha(x_s/a_0)) 
\end{eqnarray}
which at $x_s = R_0$  results in
\begin{eqnarray}
  R(\alpha, x_s)  < R(0, x_s)  <\epsilon &=& 2e^{-\frac{R_0}{a_0}} 
\end{eqnarray}
In practice this  yields a value  $\epsilon \approx 4.0 \times 10^{-3}$ for Ca$^{40}$ and $\epsilon \approx  5.0 \times 10^{-5}$ for Pb$^{208}$ respectively, 
a contribution, that is in fact  negligible compared to the exact value which is of order 1. 
Now with (\ref{qfirstcc}) we obtain for ($\ref{qq22}$)
\begin{eqnarray}
\partial_x &&\frac{1}{\Gamma(1-\alpha)}\int_0^\infty \!\!\!\!dh \frac{1}{h^\alpha}\rho(x+h) \quad |_{x=xs} 
\approx \nonumber \\
&&\partial_{x_s} \frac{1}{\Gamma(1-\alpha)}\int_0^{x_s} \!\!\!\!dh 
\frac{1}{h^\alpha}
(2 \rho(x_s) - \rho(x_s-h)) \approx\\
&& 2 \rho(x_s) \partial_{x_s} \frac{1}{\Gamma(1-\alpha)}\int_0^{x_s} \!\!\!\!dh \frac{1}{h^\alpha}
-\partial_{x_s}\frac{1}{\Gamma(1-\alpha)} \int_0^{x_s} \!\!\!\!dh \frac{1}{h^\alpha} \rho(x_s-h) 
\approx 
\nonumber \\
&& \\
&& 2 \rho(x_s)  \frac{x_s^{-\alpha}}{\Gamma(1-\alpha)}
-\partial_{x_s}\frac{1}{\Gamma(1-\alpha)} \int_0^{x_s} \!\!\!\!dh \frac{1}{h^\alpha} \rho(x_s-h) \nonumber \\
&&
\label{rieinvert}
\end{eqnarray}
The last term in (\ref{rieinvert}) is nothing else but the Riemann definition of a fractional derivative at $x_s$ \cite{rie47}:
\begin{eqnarray}
\label{riett}
_R\partial^\alpha_{x} \rho(x) = \partial_{x}\frac{1}{\Gamma(1-\alpha)} \int_0^{x} \!\!\!\!dh \frac{1}{h^\alpha} \rho(x-h) 
\end{eqnarray}
With the property that the integral converges for  $\rho(x) = x^\beta$, even for $\beta=0$:   
\begin{eqnarray}
_R\partial^\alpha_{x} x^\beta &=& \frac{\Gamma(1+\alpha)}{\Gamma(1+\alpha-\beta)} x^{\beta -\alpha}  , \qquad \beta \ge 0
\end{eqnarray}
%--------------- figure frc
\begin{figure}
\begin{center}
\includegraphics[width=\textwidth]{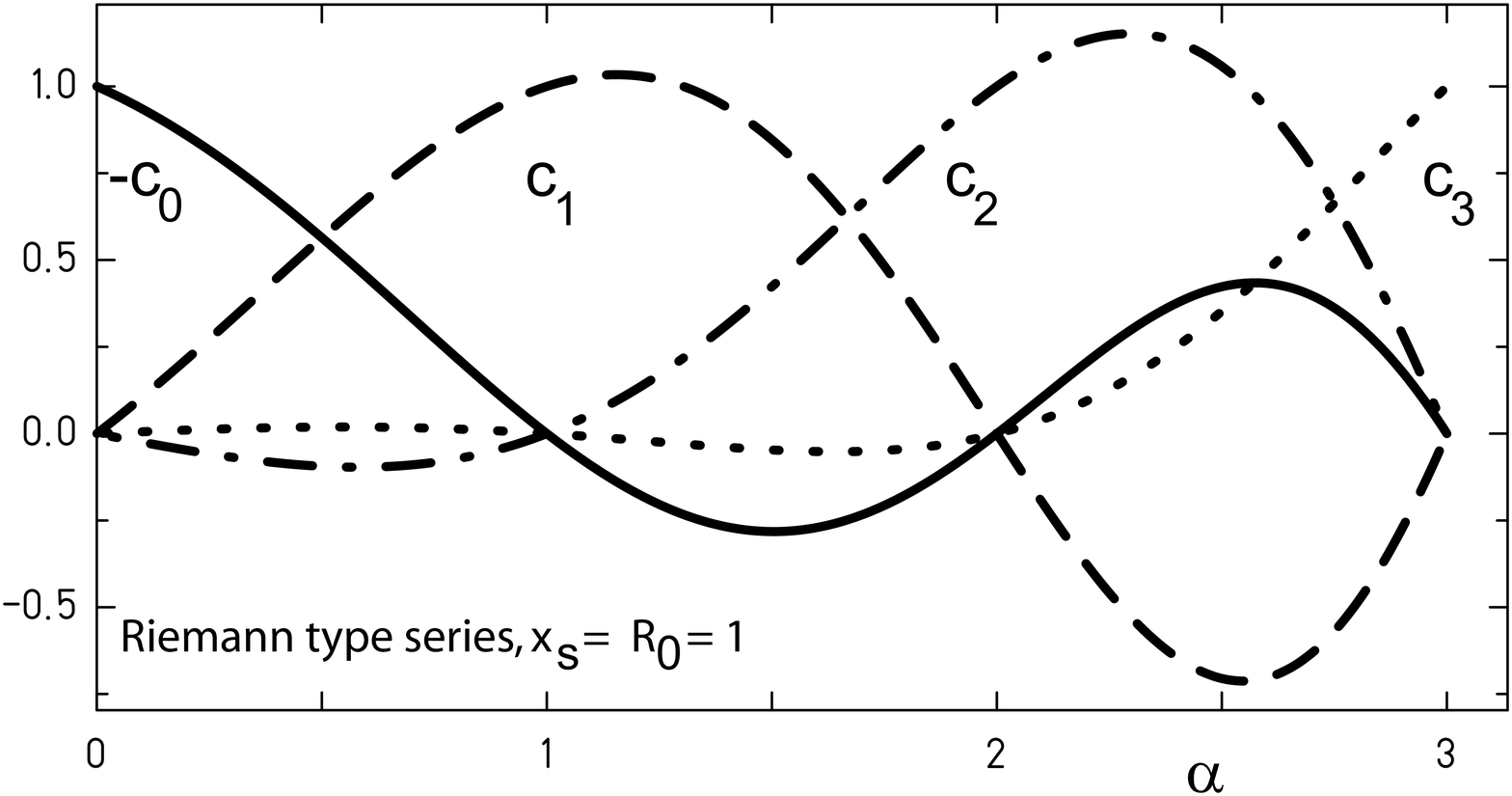}\\
\caption{
\label{seriesriemann}
{
For $^{208}\textrm{Pb}$ coefficients $c_i$ as a function of $\alpha$ from the series expansion  of the fractional Woods-Saxon potential $ \rho^{(\alpha)}_{WS}(r)$ from (\ref{dderivfrac2})  in terms of integer derivatives of order $j$ at $\tilde{x}_s=1$, see (\ref{tildex}). Signs are adjusted such that $c_j(\alpha)=1$ for $\alpha=j$. 
The relevant region $0 < \alpha < 1$ may be directly compared with the classical linear ansatz e.g. (\ref{linearAlpha})   
or (\ref{parmsBec69}). 
} }
\end{center}
\end{figure}
%-----------------------------------------------------------------------

The Leibniz product rule then follows as \cite{her18}
\begin{eqnarray}
_R\partial^\alpha_{x} \rho(x)  |_{x=xs}
&=&
\label{rdxgeneral}
 x^{-\alpha}
\sum_{j=0}^\infty 
% ssssss s t a r t binomial
\left(\begin{array}{c}
\alpha \\ j \\
\end{array} \right)
% sssssss  e n d   binomial
{ 1  \over \Gamma(1-\alpha+j)} x^j \partial_x^j \rho(x) \,\quad |_{x=xs}
\end{eqnarray}
Thus we  finally obtain 
\begin{eqnarray}
\partial_x &&\frac{1}{\Gamma(1-\alpha)}\int_0^\infty \!\!\!\!dh \frac{1}{h^\alpha}\rho(x+h) |_{x_s} 
 \nonumber \\
&& \approx 2 \rho^{(0)}(x_s)  \frac{x_s^{-\alpha}}{\Gamma(1-\alpha)}
-
 x_s^{-\alpha}
\sum_{j=0}^\infty 
% ssssss s t a r t binomial
\left(\begin{array}{c}
\alpha \\ j \\
\end{array} \right)
% sssssss  e n d   binomial
{ 1  \over \Gamma(1-\alpha+j)} x_s^j \rho^{(j)}(x_s) 
 \nonumber \\
&& = \rho^{(0)}(x_s)  \frac{x_s^{-\alpha}}{\Gamma(1-\alpha)}
-
 x_s^{-\alpha}
\sum_{j=1}^\infty 
% ssssss s t a r t binomial
\left(\begin{array}{c}
\alpha \\ j \\
\end{array} \right)
% sssssss  e n d   binomial
{ 1  \over \Gamma(1-\alpha+j)} x_s^j \rho^{(j)}(x_s) 
\end{eqnarray}
which is the series expansion of our Liouville type fractional derivative in terms of ordinary integer derivatives
$\rho^{(n)}(x_s)$ of order $n$ at $x=x_s = R_0$, with accuracy given by $R(\alpha,x_s)< \epsilon$.
In order to compare the different coefficients we perform a scaling transformation of the form
\begin{equation}
\label{tildex}
\tilde{x}_s = x_s / R_0
\end{equation}
and set $\tilde{x}_s=1$. An approximate analytic expression for the coefficients $c_j$
follows as
\begin{eqnarray}
 \rho^{(\alpha)}(\tilde{x}) |_{\tilde{x} = 1} & \approx & 
    \quad \frac{1}{\Gamma(1-\alpha)} \rho^{(0)}(\tilde{x}) |_{\tilde{x} = 1}  
   - \frac{\alpha}{\Gamma(2-\alpha)} \rho^{(1)}(\tilde{x}) |_{\tilde{x} = 1}\nonumber \\
   && \qquad \qquad  
 -
\sum_{j=2}^\infty 
% ssssss s t a r t binomial
\left(\begin{array}{c}
\alpha \\ j \\
\end{array} \right)
% sssssss  e n d   binomial
{ 1  \over \Gamma(1-\alpha+j)}  \rho^{(j)}(\tilde{x})  |_{\tilde{x} = 1}\\
& =&  \sum_{j=0}^\infty c_j  \rho^{(j)}(\tilde{x})  |_{\tilde{x} = 1}
\end{eqnarray}
We get  an analytic expression for the coefficients
\begin{equation}
\label{fft4}
c_j = 
\left\{
\begin{array}{ll}
+1/{ \Gamma(1-\alpha)}                 &\quad j=0 \\
-
% ssssss s t a r t binomial
\left(\begin{array}{c}
\alpha \\ j \\
\end{array} \right)
% sssssss  e n d   binomial
/
\Gamma(1-\alpha+j) 
                &\quad j>0
\end{array}
\right.
\end{equation}
For the lowest coefficients we explicity obtain:
\begin{equation}
\label{fft4}
c_j = 
\left\{
\begin{array}{ll}
+1/{ \Gamma(1-\alpha)}                 &\quad j=0 \\
-\alpha/{ \Gamma(2-\alpha)}                &\quad j=1\\
+\frac{1}{2}\alpha (1-\alpha)/{ \Gamma(3-\alpha)}                &\quad j=2\\
-\frac{1}{6}\alpha (1-\alpha) (2-\alpha)/{ \Gamma(4-\alpha)}                &\quad j=3
\end{array}
\right.
\end{equation}
%--------------- figure coefficients
\begin{figure}
\begin{center}
\includegraphics[width=\textwidth]{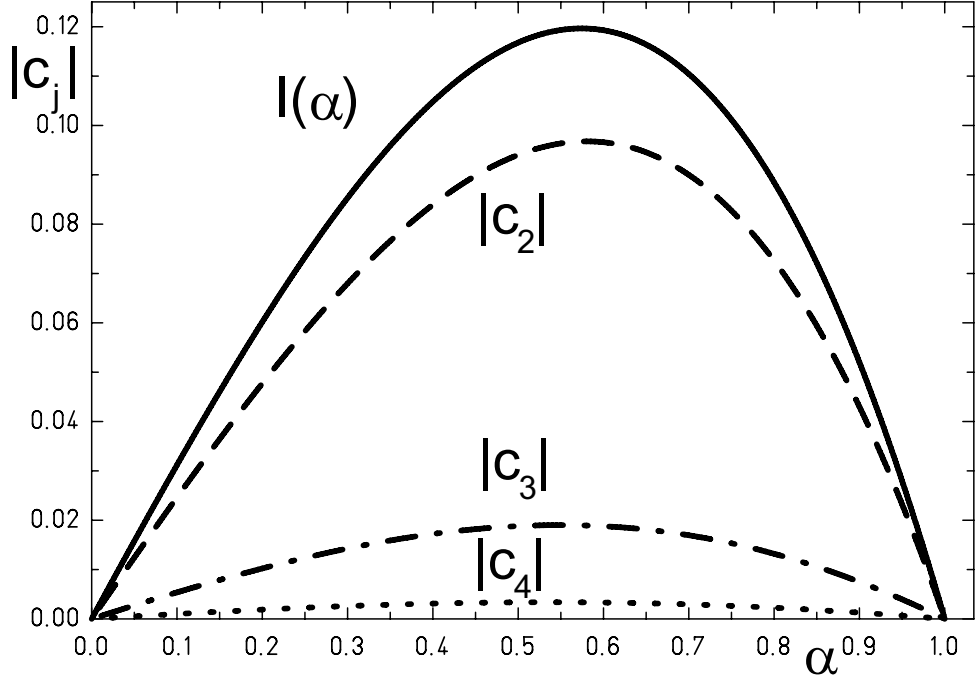}\\
\caption{
\label{cj}
{
Dashed lines show coefficients $c_j$ with $j \geq 2$  as a function of $\alpha$ from (\ref{fft4}) and 
the solid line represents the residue $I(x_s, \alpha)$ from  (\ref{riettinf}). 
} }
\end{center}
\end{figure}
In figure \ref{seriesriemann} we have plotted these coefficients.\\

Since the Riemann fractional derivative (\ref{riett}) applied to the exponential is given by:
\begin{eqnarray}
\label{riett2}
_R\partial^\alpha_{r} e^{-k r} 
&=&r^\alpha E_{1, 1-\alpha}(-k r)  \\ 
&=& (-k)^\alpha e^{-k r} (1 - \frac{\alpha \Gamma(-\alpha, -k r)}{\Gamma(1-\alpha)})\\
&=& (-k)^\alpha e^{-k r} Q(-\alpha, 0, -k r)
 \qquad \qquad  k, r \ge  0
\end{eqnarray}
where  $E_{\alpha, \beta}(z)$,  $\Gamma(a, z)$ and  $Q(a, z_0, z_1)$   are the generalized Mittag-Leffler- \cite{mit03, wim05}, the incomplete $\Gamma$- and the  generalized regularized incomplete  $\Gamma$-function \cite{mathematica}, we may  
easily obtain an analytical estimate for the influence $I(\alpha)$ of the higher order derivative contributions. 
We calculate the sum of the absolute values of the coefficients $c_j$:
\begin{eqnarray}
\label{riettinf}
I(x_s, \alpha) &=&
 \sum_{j=2}^\infty 
% ssssss s t a r t binomial
\mid
\left(\begin{array}{c}
\alpha \\ j \\
\end{array} \right) 
% sssssss  e n d   binomial
{ 1  \over \Gamma(1-\alpha+j)} 
\mid
\\
&=&
 \sum_{j=0}^\infty 
(-1)^j
% ssssss s t a r t binomial
\left(\begin{array}{c}
\alpha \\ j \\
\end{array} \right)
% sssssss  e n d   binomial
{ 1  \over \Gamma(1-\alpha+j)} 
- c_0 - c_1  \\
&=&
 (-1)^\alpha Q(-\alpha, 0, -1)
+\frac{1}{ \Gamma(1-\alpha)}   -\frac{\alpha}{ \Gamma(2-\alpha)}    
\end{eqnarray}
Within the relevant region $0 \leq \alpha\leq 1$ this function is extremal at $I(\alpha = 0.58) \approx 0.12$, while
$\mid c_0\mid + \mid c_1\mid \approx 1$. Therefore we obtain the result, that the contribution  of higher order integer derivatives  beyond $j=1$ is about $10\%$ of the $j=0,1$ part. What we have not taken into account are 
higher order function derivatives $f^{(j>1)}$  and their influence on the total fractional derivative $f^{(\alpha)}$.
In addition, our derivation is restricted to the close vicinity of $R_0$ which once again shows the limitations of a local approach.

Therefore we may deduce, that in the case of classical Woods-Saxon type functions, which fulfil the 
requirements  (\ref{qfirstcc}) and (\ref{dddsecond})  the classical approach to  generate the absorption part of the optical potential as a superposition of the original function and its first derivative may be considered as the lowest order local approximation
to an a priori nonlocal problem, the smooth transition from surface to volume absorption. 

The fractional derivative approach automatically implies a nonlocal view to the same problem and offers a consistent solution without previously made simplifications, which were necessary for the classical approach. 

It is a special property of the Woods-Saxon type functions, that the practical differences of both approaches turn out to be small 
in the cases considered so far.

\section{Conclusion}
The optical model plays a fundamental role in interpreting  scattering data in nuclear and particle physics.  In order to achieve conformity  with experimental findings, the absorption process may be understood assuming a smooth transition from surface to volume absorption with increasing energy of the incident particle. 

In this paper we proposed a new  and as we think only appropriate treatment of  this problem. 

Based on the observation, 
that a surface may be considered as a more or less drastic change of a given density and the fact, that such a change may be described using methods of vector calculus we introduced a fractional gradient definition based on methods developed in fractional calculus, which allows to determine the required smooth transition
from volume to surface potentials.

We derived closed form solutions for the practically important cases of Woods-Saxon and Woods-Saxon-plus-damped-oscillation functions in terms of higher order transcendental functions namely polylogarithms.

We applied these new fractional potentials to macroscopic and semi-microscopic nonlocal  models and found, that nonlocal effects, which are a natural property of the fractional approach, are particularly well  reproduced.

We finally showed, that the hitherto accepted classical solution, which is a simple superposition of a Woods-Saxon function and its derivative may be considered as the lowest order local approximation to the full nonlocal problem.

We have demonstrated that a complete solution of this problem can be  formulated adequately within the framework of fractional calculus.  

This paper is also a plea for optimism and a lesson how progress in physics evolves as well \cite{dir84, hos18}:

More than 60 years ago,
a first solution for optical absorption potentials was proposed, which since then has been used to categorize and interpret experimental data. Since then, this problem was considered as solved and consequently there was no demand for a more sophisticated solution nor a different vista.

Now a new point of view  based on  fractional calculus is leading 
to new insights and surprising interrelations on this classical field of research. 

It should be emphasized, that the presented fractional approach is not just one more method
to describe a smooth transition from surface to volume dependence, it is the only one 
up to now which fulfils   the claim for an appropriate treatment of the problem and of course, may be and will be  applied
to other research fields as well.

The presented results encourage further research in this field. A necessary next step will be  to use the presented fractional potentials to serve as a basis for 
the interpretation of experimental cross-sections and also an extension of the formalism to 
projectile/target combinations beyond nucleon/nucleus combinations. 

\section{Acknowledgments}
This article is the elaborated version of a talk prepared for the workshop
\emph{Fractional Differential Equations}  planned at the Isaac Newton Institute for Mathematical Sciences (INI), Cambridge UK from 4 January 2021 to 30 April 2021, but cancelled due to corona pandemic. 

We thank A. Friedrich  for  support and useful discussions. Numerical calculations were  partly performed  on the $\pi$-dron-cluster  at the
HPC testing facilities at gigahedron, Germany.

%
% Create the reference section using BibTeX:
\newpage
\section{Bibliography}
 \newcommand{\ea}{{\it et al }}


\begin{thebibliography}{}

\bibitem[{Abbot \ea (2017)}]{abb17}  Abbot, LIGO Scientific Collaboration and Virgo Collaboration (2017).
\emph{GW170817: Observation of gravitational waves from a binary neutron star inspiral }
Phys. Rev. Lett. {\bf 119}, 161101,
%\url{doi:10.1103/PhysRevLett.119.161101} 

\bibitem[{Aleksandrov \ea (2022)}]{ale22} 
Aleksandrov, I. A. Di Piazza, A. and Plunien, G. and Shabaev, V. M. (1977).
\emph{Stimulated vacuum emission and photon absorption in strong electromagnetic fields}
Phys. Rev. D.  {\bf 105}, 116005, 13pp.
%\url{doi:10.1103/PhysRevD.105.116005} 


\bibitem[{Altarelli and Parisi (1977)}]{alt77}  Altarelli, G. and Parisi, G. (1977).
\emph{Asymptotic freedom in parton language }
Nucl. Phys.  {\bf 126}, 298--318,
%\url{doi:10.1016/0550-3213(77)90384-4} 

\bibitem[{Bacon  (1267)}]{bac67} 
Bacon, R.  (1267) 
\emph{Opus majus}
Translated by Robert Belle Burke, Cambridge Library Collection - Physical Sciences, Cambridge University Press  (2010)

\bibitem[{Bauge \ea (1998)}]{bau98}  Bauge, E., Delaroche, J.~P. and Girod, M. (1998).
\emph{Semimicroscopic nucleon-nucleus spherical optical model for nuclei
with $A > 40$  at energies up to 200 MeV }
Phys. Rev. C {\bf 58}, 1118--1145,
%\url{doi:10.1103/PhysRevC.58.1118} 

\bibitem[{Bauge \ea (2001)}]{bau01a}  Bauge, E., Delaroche, J.~P. and Girod, M. (2001).
\emph{Lane-consistent, semi microscopic nucleon-nucleus optical model }
Phys. Rev. C {\bf 63}, 024607,
%\url{doi:10.1103/PhysRevC.63.024607} 

\bibitem[{Bauge (2001)}]{bau01b}  Bauge, E., Delaroche, J.~P. and Girod, M. (2001).
\emph{The MOM semi microscopic optical model potential program }
Phys. Rev. C {\bf 63}, 024607,
%\url{doi:10.1103/PhysRevC.63.024607} 

\bibitem[{Becchetti and Greenlees (1969)}]{bec69}   Becchetti, F.~D. and Greenlees, G.~W.  (1969).
\emph{Nucleon-nucleus optical-model parameters, $A > 40$, $E < 50$ MeV }
Phys. Rev.  {\bf 182}, 1190--1209,
%\url{doi:10.1103/PhysRev.182.1190} 

\bibitem[{Becquerel (1896)}]{bec96}   
Becquerel, H.  (1896).
\emph{Sur les proprieties differentes des radiations invisibles emises par les sels d'uranium }
Compt. Rendus. T. {\bf 122}, 689--694,

\bibitem[{Berger \ea (1991)}]{ber91}   
Berger, J.~F., Girod, M. and Gogny, D.~M.  (1991).
\emph{Time-dependent quantum collective dynamics applied to nuclear fission }
Comp. Phys. Comm.  {\bf 63}, 365--374,
%\url{doi:0010-4655(91)90263-K} 

\bibitem[{Bethe (1940)}]{bet40}   
Bethe, H.~A.  (1940).
\emph{A continuum theory of the compound nucleus}
Phys. Rev.  {\bf 57}, 1125--1144,
%\url{doi:10.1103/PhysRev.57.1125} 

\bibitem[{Blakemore  (1982)}]{bla82}   
Blakemore, J.~S.  (1982).
\emph{Approximations for Fermi-Dirac integrals, especially the function $F_{1/2}(\eta)$
used to describe electron density in a semiconductor}
Solid State Electronics  {\bf 25}, 1067--1076,
%\url{doi:10.1016/0038-1101(82)90143-5} 




\bibitem[{Bragg and Kleeman (1904)}]{bra04}   
Bragg, W.~H. and Kleeman, R.   (1904).
\emph{On the ionization curves of radium}
Philos. Mag.   {\bf 8:48}, 726--738,
%\url{doi:10.1080/14786440409463246} 

\bibitem[{Bragg and Kleeman (1905)}]{bra05}   
Bragg, W.~H. and Kleeman, R.   (1905).
\emph{On the $\alpha$ particles of radium, and their loss
of range in passing through various atoms and molecules}
Philos. Mag.   {\bf 10:57}, 318--340,
%\url{doi:10.1080/14786440509463378} 

\bibitem[{Capote \ea (2009)}]{cap09}   
Capote,  R., 
 Herman, M., 
Obložinský, P., 
Young, P.~G.,
Goriely, S.,
Belgya, T.,
Ignatyuk, A.~V.,
Koning, A.~J.,
Hilaire, S.,
Plujko, V.~A.,
Avrigeanu, M.,
Bersillon, O.,
Chadwick, M.~B.,
Fukahori, T.,
Ge, Z.,
Han, Y.,
Kailas, S.,
Kopecky, J.,
Maslov, V.~M.,
Reffo, G.,
Sin, M.,
Soukhovitskii, E.~Sh. and Talou, P.
  (2009).
\emph{RIPL - Reference input parameter library for calculations of nuclear reactions and nuclear data
evaluations}
Nucl. Data Sheets \textbf{110}  3107--3214 , 
Special Issue on Nuclear Reaction Data,
%\url{doi:10.1016/j.nds.2009.10.004} 

\bibitem[{Chaudhry and Qadir (2007)}]{cha07}   
Chaudhry, M.~A. and Qadir, A.    (2007).
\emph{Operator representation of Fermi-Dirac and Bose-Einstein
integral functions with applications}
Hindawi Publishing Corporation
International Journal of Mathematics and Mathematical Sciences
article ID 80515, 9 pp.
%\url{doi:10.1155/2007/805} 

\bibitem[{Costin and Garoufalidis (2007)}]{cos07}   
Costin, O. and Garoufalidis, S.   (2007).
\emph{Resurgence of the fractional polylogarithms}
arXiv:math/0701743v4 [math.CA]
%\url{doi:10.48550/arXiv.math/0701743} 

\bibitem[{Decharg\`e and Gogny (1980)}]{dec80}  
Decharg\`e, J.  and Gogny, D.  (1980).
\emph{Hartree-Fock-Bogolyubov calculations with the D1 effective interaction on spherical nuclei }
Phys. Rev. C {\bf 21}, 1568--1593,
%\url{doi:10.1103/PhysRevC.21.1568} 



\bibitem[{Dingle  (1957)}]{din57}
Dingle, R.~B.  (1957)   
\emph{The Fermi-Dirac integrals $\mathcal{F}_p (\eta ) = (p!)^{ - 1} \int\limits_0^\infty  {\varepsilon ^p (e^{\varepsilon  - \eta }  + 1} )^{ - 1} d\varepsilon $}
Applied Scientific Research  \textbf{6} 225--239,
%\url{doi:10.1007/BF02920379}

\bibitem[{Dirac  (1984)}]{dir84} 
Dirac, P.~A.~M.  (1984)
\emph{The requirements of fundamental physical theory}
Eur. J. Phys.  \textbf{5} 65--67,
%\url{doi:10.1088/0143-0807/5/2/001}



\bibitem[{Ehrenfest  (1927)}]{ehr27}
Ehrenfest, P.  (1927)   
\emph{Bemerkung über die angen\"aherte G\"ultigkeit der klassischen Mechanik innerhalb der Quantenmechanik}
Z. Phys.  \textbf{45} 455--457,
%\url{doi:10.1007/BF01329203}

\bibitem[{Fermi  (1934)}]{fer34}
Fermi, E.  (1934)   
\emph{Possible production of elements of atomic number higher than 92}
Nature \textbf{133} 898--899,
%\url{doi:10.1038/133898a0}


\bibitem[{Fermi  (1954)}]{fer54}
Fermi, L.  (1954)   
\emph{Atoms in the family: My life with Enrico Fermi},
University of Chicago Press,
ISBN 0-88318-524-5


\bibitem[{Flerov and Petrzhak  (1940)}]{fle40} 
Flerov, G.~N. and Petrzhak, K.~A.   (1940)
\emph{Spontaneous fission of uranium}
J. Phys. \textbf{3}  275--280


\bibitem[{Geiger and Marsden (1909)}]{gei09}  Geiger, H. and Marsden, E. (1909).
\emph{On a diffuse reflection of the $\alpha$-particles}
 Proc. R. Soc. Lond. A {\bf 82}, 495--500,
%\url{doi:10.1098/rspa.1909.0054} 


\bibitem[{Goldschmidt \ea (2015)}]{gol15}  Goldschmidt, A., Qiu, Z., Shen, C and Heinz, U.  (2015).
\emph{Collision geometry and flow in uranium + uranium collisions }
Phys. Rev. C {\bf 92}, 044903,
%\url{doi:10.1103/PhysRevC.92.044903} 

\bibitem[{Gomes (1959)}]{gom59} 
 Gomes, L.   (1959).
\emph{Imaginary part of the optical potential }
Phys. Rev. C {\bf 116}, 1226--1229,
%\url{doi:10.1103/PhysRev.116.1226} 




\bibitem[{Gonzales and Woods (2018)}]{gon18}  Gonzales, R.~C. and Woods, R.~E. (2018).
\emph{Digital image processing. 4$^{th}$ edition}
Pearson, Harlow, England
%\url{doi:10.1098/rspa.1909.0054} 


\bibitem[{Hahn and Stra\ss mann  (1939)}]{hah39} 
Hahn,  O. and Stra\ss mann, F.  (1939)
\emph{\"Uber den Nachweis und das Verhalten der bei der Bestrahlung des Urans mittels Neutronen entstehenden Erdalkalimetalle}
Die Naturwissenschaften \textbf{27} 11--15,
%\url{doi:10.1007/BF01488241}

\bibitem[{Herrmann (2018)}]{her18}  
Herrmann, R. (2018) 
\emph{Fractional calculus - an introduction for physicists}
3rd ed.,
World Scientific Publ., Singapore


\bibitem[{Hilfer (2000)}]{hil00}  
Hilfer, R.  (2000)
\emph{Applications of fractional calculus in physics}
World Scientific Publ., Singapore
%\url{doi:10.1142/3779}

\bibitem[{Hodgson (1967)}]{hod67}  Hodgson, P.~E. (1967).
\emph{The optical model of the nucleon-nucleus interaction}
 Annu. Rev. Nucl. Sci. {\bf 17}, 1--32,
%\url{doi:10.1146/annurev.ns.17.120167.000245} 

\bibitem[{Hofstatter (1956)}]{hof56}  Hofstatter, R. (1956).
\emph{Electron scattering and nuclear structure}
 Rev. Mod. Phys. {\bf 28}, 214--254,
%\url{doi:10.1103/RevModPhys.28.214}

\bibitem[{Hossenfelder (2018)}]{hos18}  Hossenfelder, S. (2018).
\emph{Lost in math: How beauty leads physics astray}
Basic Books; Illustrated edition (12 Jun. 2018)


\bibitem[{Joliot and Curie (1934)}]{jol34} 
Joliot, F. and Curie, I. (1934).
\emph{Artificial production of a new kind of radio-element}
Nature {\bf 133}, 201--202,
%\url{10.1038/133201a0} 


\bibitem[{Jones (1963)}]{jon63}  Jones, P.~B. (1963).
\emph{The optical model in nuclear and particle physics}
Interscience Tracts on Physics and Astronomy, {\bf 14},
Wiley and Sons, New York, London


\bibitem[{Koning and Delaroche (2003)}]{kon03}  
Koning, A.~J. and  Delaroche, J-P. (2003).
\emph{Local and global nucleon optical models from 1 keV to 200 MeV}
 Nucl. Phys. {\bf 713}, 231--310,
%\url{doi:10.1016/S0375-9474(02)01321-0} 

\bibitem[{Koning \ea (2014)}]{kon14}  
Koning, A.~J., Rochman, D.  and  van der Merk, S.~C. (2014).
\emph{Extension of TALYS  to 1 GeV}
 Nucl. Data Sheets  {\bf 118}, 187--190,
%\url{doi:10.1016/j.nds.2014.04.033} 


\bibitem[{Liouville (1832)}]{lio32} Liouville, J.  (1832)
\emph{Sur le calcul des diff\'erentielles \`a indices quelconques}
 J. \'Ecole Polytechnique  \textbf{13} 1--162,
 
 
\bibitem[{Mainardi (2010)}]{mai10}
 Mainardi, F.   (2010)
  \emph{Fractional calculus and waves in linear viscoelasticity: An introduction to mathematical models}
World Scientific Publ., Singapore


 
 \bibitem[{Meitner and Fritsch (1939)}]{mei39}
 Meitner, L. and Fritsch, O.~R.  (1939)
\emph{Disintegration of uranium by neutrons: A new type of nuclear reaction}
 Nature  \textbf{143}  239--240,
%\url{doi:10.1038/143239a0}

 \bibitem[{Mellin (1897)}]{mel97}
 Mellin, H.  (1897)
\emph{\"Uber hypergeometrische Reihen h\"oherer Ordnungen}
ex Officiina typographica Societatis litterariae fennicae \textbf{23}(7)  10pp,
Helsingfors, Finland

\bibitem[{Miller and Ross  (1993)}]{mil93}  
Miller, K.  and  Ross, B.  (1993)
  \emph{An introduction to fractional calculus and fractional differential equations}
 Wiley, New York.
 
 \bibitem[{Millev (1996)}]{mil96} 
Millev, Y.  (1903) 
\emph{Bose - Einstein integrals and domain morphology in ultrathin ferromagnetic films with perpendicular magnetization}
J. Phys. Condensed Matter \textbf{8}  3671--3676, 
%\url{10.1088/0953-8984/8/20/013}


\bibitem[{Mittag-Leffler (1903)}]{mit03} 
Mittag-Leffler, M.~G.  (1903) 
\emph{Sur la nouvelle function $E_{\alpha} (x)$}
 Comptes Rendus Acad. Sci. Paris \textbf{137} 554--558.


\bibitem[{Mueller \ea (2011)}]{mue11}
%  author = {Mueller, J. M. and Charity, R. J. and Shane, R. and Sobotka, L. G. and Waldecker, S. J. and Dickhoff, W. H. and Crowell, A. S. and Esterline, J. H. and Fallin, B. and Howell, C. R. and Westerfeldt, C. and Youngs, M. and Crowe, B. J. and Pedroni, R. S.},
 Mueller, J.~M. \ea  (2011)
\emph{Asymmetry dependence of nucleon correlations in spherical nuclei extracted from a dispersive-optical-model analysis}
 Phys. Rev. C  \textbf{83}  064605, 32pp., 
%\url{10.1103/PhysRevC.83.064605}

\bibitem[NIST (2012)]{nis12} 
National Institute of Standards and Technology  (NIST)  (2012)
\emph{Chemistry webBook, NIST standard reference database number 69} 
%\url{http://webbook.nist.gov/chemistry/}



\bibitem[{Oganessian \ea  (2004)}]{oga04}
 Oganessian, Yu.~Ts. \ea
 (2004)
\emph{Measurements of cross sections for the fusion-evaporation reactions Pu-244 (Ca-48,xn) (292-x)114 and Cm-245 (Ca-48,xn) (293-x)116}
Phys. Rev. C \textbf{69}   054607,
%\url{doi:10.1103/PhysRevC.69.054607}
 
 \bibitem[{Oldham and Spanier (1974)}]{old74} 
Oldham, K.~B. and Spanier, J.  (1974)
 \emph{The fractional calculus}
Academic Press, New York

 \bibitem[{Oppenheimer and Volkoff (1939)}]{opp39}
Oppenheimer, J.~ R. and Volkoff, G.~M.  (1939)
\emph{On massive neutron cores}
Phys. Rev.  \textbf{55}  374--381,
%\url{doi:10.1103/PhysRev.55.374}

\bibitem[{Ortigueira (2011)}]{ort11} 
Ortigueira, M.~D.  (2011)
\emph{Fractional calculus for scientists and engineers}
Springer, Berlin, Heidelberg, New York


\bibitem[{Ostrofsky \ea (1936)}]{ost36}  
Ostrofsky, M, Breit, G. and Johnson, D.~P.  (1936).
\emph{The excitation function of lithium under proton bombardment }
Phys. Rev.  {\bf 49},22--34,
%\url{doi:10.1103/PhysRev.49.22} 

\bibitem[{Pauli  (1927)}]{pau27}  
Pauli, W.  (1927).
\emph{\"Uber Gasentartung und Paramagnetismus }
Z. Phys. A  {\bf 41}, 81--102,
%\url{doi:10.1007/BF01391920} 

\bibitem[{Perey and Buck (1962)}]{per62}  
Perey, F.  and Buck, B.   (1962).
\emph{A nonlocal potential model for the scattering of neutrons by nuclei }
Nucl. Phys.  {\bf 32}, 353--380,
%\url{doi:10.1016/0029-5582(62)90345-0} 

\bibitem[{Podlubny (1999)}]{pod99} 
Podlubny, I.  (1999)
 \emph{Fractional differential equations}
Academic Press, New York

\bibitem[{Rafelski and  M\"uller (1982)}]{raf82}  Rafelski, J.  and M\"uller (1982).
\emph{Strangeness production in the quark-gluon plasma }
Phys. Rev. Lett.  {\bf 48},1066--1069,
%\url{doi:10.1103/PhysRevLett.48.1066} 

\bibitem[{Rafelski and  M\"uller (1986)}]{raf86}  Rafelski, J.  and M\"uller (1986).
\emph{Erratum: strangeness production in the quark-gluon plasma }
Phys. Rev. Lett.  {\bf 56}, 2334,
%\url{doi:10.1103/PhysRevLett.56.2334} 

\bibitem[{Rapaport \ea (1986)}]{rap79}  
Rapaport, J. , Kulkarni, V.  and Finlay, R.~W. (1979).
\emph{  A global optical-model analysis of neutron elastic scattering data }
Nucl. Phys.  {\bf A330}, 15--28,
%\url{doi:10.1016/0375-9474(79)90533-5} 


\bibitem[{Reinhardt and Greiner (1977)}]{rei77}  Reinhardt, J.  and Greiner, W. (1977).
\emph{Quantum electrodynamics of strong fields}
Rep. Prog. Phys.  {\bf 40}, 219--295,
%\url{doi:10.1088/0034-4885/40/3/001} 


\bibitem[{Reinhardt \ea (1981)}]{rei81}  
Reinhardt, J. , M\"uller, B.  and Greiner, W. (1981).
\emph{Theory of positron production in heavy-ion collisions}
Phys. Rev. A {\bf 24}, 103--128,
%\url{doi:10.1103/PhysRevA.24.103} 

\bibitem[{Riemann (1847)}]{rie47}  
Riemann, B.  (1847)
 \emph{Versuch einer allgemeinen Auffassung der Integration
und Differentiation} in:  Weber, H. and Dedekind, R.   (Eds.)  (1892)
\emph{Bernhard Riemann's gesammelte mathematische Werke und wissenschaftlicher Nachlass}  
Teubner, Leipzig, reprinted in
\emph{Collected works of Bernhard Riemann}  
Dover Publications  (1953) 353--366



\bibitem[{Rose and Jones (1984)}]{ros84} 
Rose, H.~J. and Jones, G.~A.  (1984) 
\emph{A new kind of natural radioactivity}
Nature \textbf{307}  245--247,
%\url{doi:10.1038/307245a0}

\bibitem[{Rutherford (1911)}]{rut11}  Rutherford, R. (1911).
\emph{The scattering of $\alpha$ and $\beta$ particles by matter and the structure of the atom }
The London, Edinburgh, and Dublin Philosophical
Magazine and Journal of Science {\bf 21}, LXXIX.,  669--688,
%\url{doi:10.1080/14786440508637080} 

\bibitem[{Rutherford (1919)}]{rut19}  Rutherford, R. (1919).
\emph{Collision of $\alpha$ particles with light atoms. IV. An anomalous effect in nitrogen }
The London, Edinburgh, and Dublin Philosophical
Magazine and Journal of Science {\bf 37}, LIV.,  581--587,
%\url{doi:10.1080/14786440608635919} 

\bibitem[{Sachdev  (1993)}]{sac93} 
Sachdev, S.  (1993).
\emph{Polylogarithm identities in a conformal field theory
in three dimensions}
 Phys. Let. B   \textbf{309}  285--288,
%\url{doi:10.1016/0370-2693(93)90935-B}

\bibitem[{Samko  \ea (1993)}]{sam93} 
Samko, S.~G., Kilbas, A.~A.  and  Marichev, O.~I.  (1993)
\emph{Fractional integrals and derivatives} Translated from the 1987 Russian original, Gordon and Breach, Yverdon


\bibitem[{Satchler and Love  (1979)}]{sat79} 
Satchler, G.~R. and Love, W.~G.  (1979).
\emph{Folding model potentials from realistic interactions for heavy-ion scattering}
 Phys. Rep.   \textbf{55}  183--254,
%\url{doi:10.1016/0370-1573(79)90081-4}

\bibitem[{Sommerfeld  (1928)}]{som28}  
Sommerfeld, A.  (1928).
\emph{Zur Elektronentheorie der Metalle auf Grund der Fermischen Statistik }
Z. Phys. A  {\bf 47}, 1--32,
%\url{doi:10.1007/BF01391052} 

 
\bibitem[{Tao (2022)}]{tao22}  
Tao, J. (2022).
\emph{Zeta annuities, fractional calculus, and polylogarithms }
available at SSRN: ssrn.com/abstract=4049283
%\url{doi:10.2139/ssrn.4049283}
 

\bibitem[{Tarasov (2008)}]{tar08}  Tarasov, V.~E.  (2008).
\emph{Fractional vector calculus and fractional Maxwell’s equations }
Annals of Physics {\bf 323}, 2756--2778
%\url{doi:10.1016/j.aop.2008.04.005}

\bibitem[{Tarasov (2016)}]{tar16}  Tarasov, V.~E.  (2016).
\emph{Leibniz rule and fractional derivatives of power functions }
J. Comput. Nonlinear Dynam., {\bf 11}, 031014, 4 pp.
%\url{doi:10.1115/1.4031364}

\bibitem[{Tarasov (2021)}]{tar21}  Tarasov, V.~E.  (2021).
\emph{General fractional vector calculus }
Mathematics, {\bf 9}, 2816
%\url{doi:10.3390/math9212816}


\bibitem[{Thomson (1904)}]{tho04}  Thomson, J.~J.  (1904).
\emph{XXIV. On the structure of the atom: an investigation of the stability and periods
of oscillation of a number of corpuscles arranged at equal intervals around the circumference of a circle; with
application of the results to the theory of atomic structure }
Philosophical Magazine Series 6, 7 {\bf 39}, 237--265
%\url{doi:10.1080/14786440409463107} 





\bibitem[{Tian \ea (2015)}]{tia15} 
 Tian, Y.,   Pang, D.~Y.  and  Ma,  Z.~Y.  (2015).
\emph{Systematic nonlocal optical model potential for nucleons }
Int. J. Mod. Phys. E,   {\bf 24} , 1550006
%\url{doi:0.1142/S0218301315500068}

\bibitem[{Tooper (2021)}]{too69}  Tooper, R.~F.  (1969).
\emph{On the equation of state of a relativistic Fermi-Dirac as at high temperatures }
APJ, {\bf 156}, 1075--1100
%\url{doi:10.1086/150036}


\bibitem[{Varner \ea (1991)}]{var91} 
Varner, R.~L.,  Thompson, W.~J, McAbee, T.~L.  Ludwig, E.~J.  and Clegg, T.~B,
 (1991)
\emph{A global nucleon optical model potential}
 Phys. Rep.   \textbf{201}  57--119,
%\url{doi:10.1016/0370-1573(91)90039-O}

\bibitem[{Walter and Guss (1986)}]{wal86} 
Walter, R.~L  and Guss, P.~P,
 (1986)
\emph{A global optical model for neutron scattering for $A > 53$ and $10 < E < 80$  MeV}
Proc. Int. Conf. on Nuclear Data for Basic and Applied Sciences, Santa Fe, N.M., U.S.A.,
Gordon and Breach, p.1079 
%\url{doi:10.1080/00337578608208670}

\bibitem[{Wilson (1946)}]{wil46} 
Wilson, R.~R.,  (1946)
\emph{Radiological use of fast protons}
Radiology   \textbf{47:5}  487--491,
%\url{doi:10.1148/47.5.487}

\bibitem[{Wiman (1905)}]{wim05} 
Wiman, A.  (1905) 
\emph{\"Uber den Fundamentalsatz in der Theorie der Funktionen $E_a (x)$}
Acta Math. \textbf{29} 191--201,
%\url{doi:10.1007/BF02403202}

\bibitem[{Wolfram (2022)}]{mathematica} 
Wolfram, S.  (2022)
\emph{Wolfram Mathematica documentation center}
%\url{http://reference.wolfram.com/mathematica/guide/Mathematica.html}.

\bibitem[{Wolt (2017)}]{wol17} 
Wolt, P.  (2017)
\emph{Not even wrong: The failure of string theory and the search for unity in physical law},
Basic Books, NY


\bibitem[{Wood (1992)}]{woo92}  Wood, D.~C. (1992).
\emph{The computation of polylogarithms}, Technical Report 15-92, Canterbury, UK: University of Kent Computing Laboratory


\bibitem[{Woods \ea (1982)}]{woo82}  Woods, C.~L., Brown, B.~A.  and Jelley, N.~A.   (1982).
\emph{A comparison of Woods-Saxon and double-folding potentials for lithium scattering from light target nuclei}
J. Phys. G.: Nucl. Phys. {\bf 8}, 1699--1719,
%\url{doi:10.1088/0305-4616/8/12/012} 

\bibitem[{Woods and Saxon (1954)}]{woo54} 
 Woods, R.~D. and Saxon, D.~S.   (1954).
\emph{Diffuse surface optical model for nucleon-nuclei scattering }
Phys. Rev. {\bf 95}, 577--578
%\url{doi:10.1103/PhysRev.95.577} 



\bibitem[{Xu \ea (2016)}]{xu16} 
Xu, C. , Yan, Y. and Shi, Z.   (2016).
\emph{Euler sums and integrals of polylogarithm functions }
J. Num. Th. {\bf 165}, 84--108
%\url{doi:10.1016/j.jnt.2016.01.025} 

\end{thebibliography}
\end{document}